\documentclass[twocolumn,backref, breaklinks, colorlinks]{aastex63}
\hypersetup{linkcolor=blue,citecolor=blue,filecolor=cyan,urlcolor=magenta}

\usepackage{tabularx}
\usepackage{url}
\usepackage{graphicx}
\usepackage[T1]{fontenc}
\usepackage{ae,aecompl}
\usepackage{booktabs}
\usepackage{natbib}
\usepackage{enumitem}
\usepackage{mathtools}
\usepackage{graphicx}
\usepackage{amssymb}
\usepackage{amsmath}
\usepackage{ulem}
\usepackage{epstopdf}
\usepackage{color}
\usepackage{graphicx}
\usepackage{xcolor}
\usepackage{needspace}
\usepackage{blindtext}
\usepackage{longtable}
\usepackage{tabu}
\usepackage[title]{appendix}

\def\unit #1{\,{\rm #1}}

\newcommand\kms{\rm \,\unit{km\,s^{-1}}}

\newcommand\cmsqi{\rm \,\unit{cm^{-2}}}

\newcommand\kev{\rm \,\unit{keV}}

\newcommand\funit{\rm \,erg\,cm^{-2}\,s^{-1}}

\newcommand\lunit{\rm \,erg \,s^{-1}}

\newcommand\xiunit{\rm \,erg\,cm\,s^{-1}}

\newcommand\lambdaedd{\lambda_{\rm \, Edd}}

\newcommand\lbol{L_{\rm \, bol}}

\newcommand\mdot{\dot{M}}

\newcommand\msol{M_{\odot}}

\newcommand\mbh{M_{\rm BH}}

\newcommand\nh{\rm N_{H}}

\newcommand\ks{\, \rm ks}

\newcommand\dc{\, \Delta\chi^2}

\newcommand\cd{\,\rm \chi^2/dof}

\newcommand\mpc{\unit{Mpc}}

\newcommand\ev{\unit{\, eV}}

\newcommand\chandra{{\it Chandra}}

\newcommand\swift{{\it Swift}}
\newcommand\xmm{{\it XMM-Newton}}

\newcommand\suzaku{{\it Suzaku}}
\newcommand\nustar{{\it NuSTAR}}

\def\aap{A\&A} \def\apjl{ApJ}  
 \def\apjs{ApJS}

 \submitjournal{ApJ}

\shorttitle{Vanishing soft excess in Mrk~590}

\begin{document}

\title{The origin of the vanishing soft X-ray excess in the changing-look Active Galactic Nucleus Mrk~590}

\author[0000-0003-4790-2653]{Ritesh Ghosh}
\affiliation{Inter-University Centre for Astronomy and Astrophysics (IUCAA), Pune, 411007, India.}

\author[0000-0003-2714-0487]{Sibasish Laha}

\affiliation{Astrophysics Science Division, NASA Goddard Space Flight Center, Greenbelt, MD 20771, USA.}
\affiliation{Center for Space Science and Technology, University of Maryland Baltimore County, 1000 Hilltop Circle, Baltimore, MD 21250, USA.}
\affiliation{Center for Research and Exploration in Space Science and Technology, NASA/GSFC, Greenbelt, Maryland 20771, USA}

\author[0000-0001-5253-3480]{Kunal Deshmukh}
\affiliation{Department of Metallurgical Engineering and Materials Science, Indian Institute of Technology Bombay, Powai, Mumbai 400076, India.}

\author[0000-0000-0000-0000]{Varun Bhalerao}
\affiliation{Department of Physics, Indian Institute of Technology Bombay, Powai, Mumbai 400076, India.}

\author[0000-0000-0000-0000]{Gulab C. Dewangan}
\affiliation{Inter-University Centre for Astronomy and Astrophysics (IUCAA), Pune, 411007, India.}

\author[0000-0000-0000-0000]{Ritaban Chatterjee}
\affiliation{Department of Physics, Presidency University, 86/1 College Street, Kolkata 700073, India.}



\correspondingauthor{Ritesh Ghosh}
\email{ritesh.ghosh1987@gmail.com, ritesh.ghosh@iucaa.in}


\begin{abstract}

We have studied the nature and origin of the soft X-ray excess detected in the interesting changing-look AGN (CLAGN) Mrk~590 using two decades of multi-wavelength observations from \xmm{}, \suzaku{}, \swift{} and \nustar{}. In the light of vanishing soft excess in this  CLAGN, we test two models, ``the warm Comptonization'' and ``the ionized disk reflection'' using extensive UV/X-ray spectral analysis. Our main findings are: (1) the soft X-ray excess emission, last observed in 2004, vanished in 2011, and never reappeared in any of the later observations, (2) we detected a significant variability ($\sim300\%$) in the observed optical-UV and power-law flux between observations with the lowest state ($\lbol = 4.4\times 10^{43}\lunit$, in 2016) and the highest state ($\lbol = 1.2\times 10^{44}\lunit$, in 2018), (3) the UV and power-law fluxes follow same temporal pattern, (4) the photon index showed a significant variation ($\Gamma=1.88^{+0.02}_{-0.08}$ and $\Gamma=1.58^{+0.02}_{-0.03}$ in 2002 and 2021 respectively) between observations, (5) no Compton hump was detected in the source spectra but a narrow Fe$K_{\alpha}$ line is present in all observations, (6) we detected a high-energy cut-off in power-law continuum ($92^{+55}_{-25}\kev$ and $60^{+10}_{-08}\kev$) with the latest \nustar{} observations, (7) the warm Comptonization model needs an additional diskbb component to describe the source UV bump, (8) there is no correlation between the Eddington rate and the soft excess as found in other changing-look AGNs. We conclude that given the spectral variability in UV/X-rays, the ionized disk reflection or the warm Comptonization models may not be adequate to describe the vanishing soft excess feature observed in Mrk~590.

\end{abstract}

\vspace{0.5cm}

\section{INTRODUCTION}

The X-ray continuum of Active Galactic Nuclei (AGNs) is mostly dominated by a power-law component arising in a hot corona via inverse Compton scattering of soft seed photons. The presence of soft X-ray excess (soft excess from here on) emission below $2\kev$ is commonly observed in the X-ray spectra of type 1 AGNs and is often used to study in detail the accretion disk/corona geometry and the physical processes that govern it. This soft excess emission was first discovered in the 1980s, \citep{1985MNRAS.217..105A, 1985ApJ...297..633S} and since then has been observed in a large fraction of AGNs over time, and using different X-ray telescopes, \citep{1986MNRAS.223P..29B,1988MNRAS.232..463T,1992A&A...255..119G,1994ApJ...435..611L,1997A&AS..126..525P, 2001ApJ...559..181P, 2004MNRAS.349L...7G,2007ApJ...671.1284D, 2010MNRAS.406.2591P, 2011MNRAS.410.1251N, 2013ApJ...777....2L, 2014MNRAS.437.2664L, 2016MNRAS.456..554G, 2018MNRAS.479.2464G, 2018A&A...609A..42P, 2019MNRAS.486.3124L, 2019ApJ...871...88G, 2020MNRAS.497.4213G,2020A&A...640A..99M, 2021ApJ...908..198G}. Characterizing the soft excess is an important tool in investigating the AGN central region that is still unresolved with the state of the art telescopes. However, the physical origin of soft excess is still debated in literature \citep{2006MNRAS.365.1067C, 2012mnras.420.1848d, 2019ApJ...871...88G, 2021ApJ...913...13X, 2021ApJ...908..198G}.

Historically, the type 1 AGNs have been favored to study this excess emission as they provide us with a direct view of the spatially unresolved central region \citep{1995PASP..107..803U} of the AGNs.  However, recent studies have detected large spectral state changes in AGNs that challenges our current understanding of Type 1 and Type 2 AGN classification. In the last couple of years, a dozen luminous ``changing-look AGNs'' (CLAGNs) \citep{2003MNRAS.342..422M} were discovered to exhibit strong, persistent changes in luminosity, accompanied by the dramatic emergence or disappearance of broad Balmer emission-line~\citep{2014ApJ...788...48S,2014ApJ...796..134D,2015ApJ...800..144L,2018ApJ...862..109Y,2019ApJ...874....8M}. For most of the sources, this changing-look behavior is considered as an intrinsic property of the central engine~\citep{2017ApJ...846L...7S,2018ApJ...866..123M,2018ApJ...864...27S,2019A&A...625A..54H} implying that ``type" is not always associated with the viewing angle of the observer. Some of the possible explanations that have been put forward by these studies are, (1) changing the inner disk radius leading to state transition \citep{2019ApJ...883...76R,2018MNRAS.480.3898N}, (2) radiation pressure instabilities in the disk \citep{2020A&A...641A.167S}, (3) tidal disruptive events (TDEs) \citep{2020ApJ...898L...1R}, (4) variation in the accretion rate \citep{2014MNRAS.438.3340E}, and (5) variable obscuration causing a switch from a Compton-thick to Compton-thin absorption in the X-ray band \citep{2002MNRAS.329L..13G,2003MNRAS.342..422M}. Hence studying the origin of soft excess in a changing-look AGN can shed light not only on the cause of changing-look but also the relation between soft excess and changing-look nature.  

Mrk~590 (also known as NGC 863) is a nearby (z = 0.0264), X-ray bright CLAGN, which has shown similar dramatic changes in amplitude of broad Balmer emission lines \citep{1993ApJ...414..552O, 2014ApJ...796..134D, 2018ApJ...866..123M,2019MNRAS.486..123R}. The source has changed from type 1.5~\citep{1977ApJ...215..733O} to type 1~\citep{1998ApJ...501...82P} and then to type $\sim 1.9-2$~\citep{2014ApJ...796..134D}. Mrk~590 also showed significant variability in luminosity at optical wavelength as the central AGN brightened by a factor of $\sim 10$ between the 1970s and 1990s, then faded by a factor of $\sim100$ between the 1990s and 2013. \cite{2014ApJ...796..134D} suggested the change in source luminosity due to drop in black hole accretion rate. Later, \cite{2018ApJ...866..123M} studied the \chandra{} and {\it Hubble} Space Telescope observations from 2014 and showed that Mrk 590 was changing its appearance again to type 1, most possibly due to episodic accretion events. \cite{2019MNRAS.486..123R} discovered that after $\sim 10$ years of absence, the optical broad emission lines of Mrk 590 have reappeared. However, the optical continuum flux was still $\sim 10$ times lower than that observed during the most luminous state in the 1990s. In 2015, \cite{2021MNRAS.502L..61Y} studied the source with very long baseline interferometry (VLBI) observations with the European VLBI Network (EVN) at 1.6 GHz and found a faint ($\sim1.7\rm mJy$) radio jet extending up to $\sim 1.4 pc$. Both parsec-scale jet and type changes in Mrk~590 were attributed to variable accretion onto the super massive black hole (SMBH). The study of X-ray spectra of Mrk~590 also revealed very interesting features~\citep{2012ApJ...759...63R}. The soft excess emission present in the \xmm{} observation in 2004 have vanished in the 2011 \suzaku{} observations while the photon index and the $2-10 \kev$ continuum flux have varied only minimally ($10\%$). The 2013 \chandra{} observation \citep{2018ApJ...866..123M} showed the source to be still in a low state, however, the presence of a weak soft excess was observed. 

This variability in the soft excess flux in a nearby, X-ray bright source such as Mrk~590, provides us with an opportunity to study in detail the origin and nature of this emission in the light of its changing-look nature. In this work, the main science goals we want to address are, (1) the origin and nature of the soft excess emission and (2) to investigate the likely cause of the type change in this CLAGN. We use multi-epoch and multi-wavelength observations of Mrk~590 available in the {\sc HEASARC} archive. We used two physically motivated models, the relativistic reflection from an ionized accretion disk and the intrinsic thermal Comptonization, to describe the soft excess emission. 

The paper is organised as follows. Section \ref{sec:obs} describes the observation and data reduction techniques. The steps taken in the spectral analysis are discussed in  Section \ref{sec:analysis}. Section \ref{sec:results} includes the main results followed by in-depth discussion in Section~\ref{sec:discussion} and finally conclusions in Section~\ref{sec:conclusion}. Throughout this paper, we assumed a cosmology with $H_{0} = 71\kms \mpc^{-1}, \Omega_{\Lambda} = 0.73$ and $\Omega_{M} = 0.27$.   \\

\section{Observation and data reduction}\label{sec:obs}

We have used multi-epoch, multi-wavelength data sets publicly available in the HEASARC archive as on January 2021. Our observations span a baseline of almost 20 years from 2002 to 2021. We have included all the available simultaneous \swift{} and \nustar{} observations, except for the one in 2019 (\nustar{} observation was heavily affected by Solar Coronal Mass Ejections). We have studied two \xmm{}, two \suzaku{} and four simultaneous \nustar{} plus \swift{} observations (See Table~\ref{Table:obs} for details). There are also three \chandra{} observations of this source available in the archive ~\citep{2007A&A...470...73L,2018ApJ...866..123M}. However, these observations have very poor signal-to-noise ratio above $7\kev$, crucial to constrain the power-law and neither have simultaneous UV flux. Hence we did not use them in our work.

\begin{table}
\centering

{\footnotesize
  \caption{The X-ray observations of Mrk~590 used in our work. \label{Table:obs}}
  \begin{tabular}{cccccccc} \hline\hline 

X-ray		& observation	&Short	&Date of obs	& Net	\\
Satellite	&id		        &id	    &		        & Exposure	\\ \hline 

\xmm{}          &0109130301	&obs1	&01-01-2002	&$11\ks$       \\
	        &0201020201	&obs2	&04-07-2004	&$113\ks$       \\

\suzaku{}	&705043010	&obs3	&23-01-2011	&$62\ks$	\\
		&705043020	&obs4	&26-01-2011	&$41\ks$	\\

\nustar{}	&60160095002	&obs5	&05-02-2016	&$22\ks$	\\
\swift{}	&00080903001	&	&05-02-2016	&$6\ks$	\\

\nustar{}	&80402610002	&obs6	&27-10-2018	&$21\ks$	\\
\swift{}	&00010949001	&	&28-10-2018	&$2\ks$	\\

\nustar{}	&80502630004	&obs7	&21-01-2020	&$50\ks$	\\
\swift{}	&00013172002	&	&21-01-2020	&$5\ks$	\\

\nustar{}	&80502630006	&obs8	&10-01-2021	&$42\ks$	\\
\swift{}	&00095662033	&	&10-01-2021	&$10\ks$	\\

\hline\hline
\end{tabular}  
}
\end{table}

\subsection{\xmm{}}

The \xmm{} observed Mrk~590 in 2002 January 01 and then in 2004 July 04. The details of the observations and the short ids are mentioned in Table~\ref{Table:obs}. Archival data from the EPIC, RGS and OM instruments are available. We preferred the EPIC-pn \citep{2001A&A...365L..18S} over MOS data due to their better signal-to-noise ratio which is critical for the broadband spectral study of our source. For both observations (obs1 and obs2), the EPIC-pn camera operated in the small-window mode. The EPIC-pn data were reprocessed with V18.0.0 of the Science Analysis Software (SAS)~\citep{2004ASPC..314..759G} using the task {\it epchain}. We created the filtered event list after screening for flaring background due to high-energy particles. Circular region of 40 arcsec, centered on the centroid of the source were used to extract the source counts whereas 40 arsec circular region, away from the source but located on the same CCD, was selected to estimate the background counts. The SAS task {\it epatplot} was used to estimate the pile-up in our observations. We found that both obs1 and obs2 are free of any pile-up. The corresponding response matrix function (RMF) and auxiliary response function (ARF) for each observations were created employing the SAS tasks {\it arfgen} and {\it rmfgen}. We used the command {\it specgroup} to group the \xmm{} spectra by a minimum of 20 counts per channel and a maximum of three resolution elements required for $\chi^{2}$ minimization technique. The task {\it omichain} was used to reduce the data from the Optical Monitor for the six active ﬁlters (V, B, U, UVW1, UVM2 and UVW2). We used the task {\it om2pha} to create the necessary files to be analysed with {\sc XSPEC}, together with the simultaneous X-ray data. We corrected the observed UV fluxes for the Galactic reddening assuming \citep{1999PASP..111...63F} reddening law with $R_{\rm v}= 3.1$. We fixed the color excess parameter of the redden component at $E(B-V)= 0.0306$  and the Galactic extinction coefficient value used was 0.257 (Mrk~590).

\subsection{\suzaku{}}

\suzaku{} started observing Mrk~590 in 2011 January 23, however, interrupted due to a Target of Opportunity trigger. The observation continued on 2011 January 26 making them two separate observations (See Table~\ref{Table:obs}. \suzaku{} has three X-ray Imaging
Spectrometers (XISs) \citep{2007PASJ...59S..23K} along with the Hard X-ray Detector (HXD) \citep{2007PASJ...59S..35T} on board that cover a broad energy band of $0.2-50\kev$. There are however no simultaneous optical-UV observations. In both observations (obs3 and obs4), the XIS data were obtained in both $3 \times 3$ and $5 \times 5$ data mode and XIS nominal position. We reprocessed the data using \suzaku{} pipeline with the screening criteria recommended in the Suzaku Data Reduction Guide. All extractions were done using HEASOFT (V6.27.2) software and the recent calibration ﬁles. For HXD/PIN, which is a non-imaging instrument, we used appropriate tuned background ﬁles provided by the Suzaku team and available at
the HEASARC website. We co-added the spectral data from the front-illuminated XIS instruments to enhance the signal-to-noise ratio. We used the tool {\it grppha} to group the XIS spectral data from both the observations to a minimum of 100 counts in each energy bin. We also grouped the HXD/PIN data to produce $\sim 60$ energy bins with more than 20 counts per bin.

\subsection{\nustar{}}

We have used four quasi-simultaneous \nustar{} and \swift{} observations  of Mrk~590 (obs5, obs6, obs7 and obs8). See Table~\ref{Table:obs} for details. These are all the currently available archival data that are free of any technical issues reported in \nustar{} Master Catalog. Year 2016 and 2021, both have two \nustar{} observations that have simultaneous \swift{} data. We selected those two \nustar{} observations, one each from 2016 and 2021, that have highest exposure in the \swift{} XRT instrument. This is essential for detecting the soft excess in the X-ray band. We reprocessed the \nustar{} FPM \citep{2013ApJ...770..103H} and \swift{} XRT \citep{2004SPIE.5165..201B} plus UVOT \citep{2005SSRv..120...95R} data. For \nustar{} we produced the cleaned event files using the standard NUPIPELINE (v2.0.0) command, part of {\sc HEASOFT} V6.28 package, and instrumental responses from \nustar{} CALDB version V20210202. For lightcurves and spectra, we used a circular extraction region of 80 arcsec centered on the source position and a 100 arcsec radius for background, respectively. The \nustar{} spectra were grouped to a minimum count of 20 per energy bin, using the command {\it grppha} in the {\sc HEASOFT} software.

\subsection{\swift{}}
The \swift{} XRT and UVOT data reprocessing and spectral extraction were carried out following the steps described in \cite{2016MNRAS.456..554G, 2018MNRAS.479.2464G}. All the \swift{} XRT spectra were grouped by a minimum of 20 counts per channel. In both observations, \swift{} UVOT observed the source in all the six filters i.e. in the optical (V, B, U) bands and the near UV (W1, M2, W2) bands. We used the {\it UVOT2PHA} tool to create the source and background spectra and used the response files provided by the \swift{} team.

\section{Data Analysis}\label{sec:analysis}

All the spectral fitting were done using the XSPEC \citep{1996aspc..101...17a} software. Uncertainties quoted on the fitted parameters reflect the 90 per cent confidence interval for one interesting parameter, corresponding to $\Delta \chi^{2} = 2.7$ \citep{1976ApJ...208..177L}. We used the solar abundances from \citet{2000ApJ...542..914W} and cross-sections from \citet{1996ApJ...465..487V}. The Galactic column density value used in our work is  $\rm N_{H} = 2.77\times 10^{20} \rm cm^{-2}$ \citep{1990ARA&A..28..215D}, modelled by {\it tbabs}, for all spectral analysis done in this work. We started with a set of phenomenological models to statistically detect the different spectral features present at each epoch and then used physically motivated models to describe the spectral evolution. We fitted the data sets for each observation separately in all our spectral analysis. The simultaneous \nustar{} and \swift{} data were fitted together.

\subsection{The phenomenological models}\label{subsec:pheno}

We began our spectral fitting of the source spectra with a set of phenomenological models. This exercise helps us to characterize the source spectra at different epoch and determine the spectral features quantitatively. For Mrk~590, we used a power-law representing the primary continuum emission, {\it diskbb} to model the soft excess, {\it zgauss} for the Fe line emission, and {\it pexrav} with a negative reflection fraction to model the Compton hump. The power-law, diskbb and Gaussian model components are made free to vary during the spectral fitting of each observation to check the variability of the continuum and discrete spectral properties of the source. We fixed all {\it pexrav} model components except for the reflection fraction. We tied the photon index and model normalization of {\it pexrav} with that of the primary continuum. The abundance of Iron and other heavier elements than He was fixed to solar values. We made the inclination of {\it pexrav} a free parameter for obs3 only. This particular observation was selected due to its longest exposure among the \suzaku{} and \nustar{} data sets. We used this best-fit inclination value and fixed it for all other observations. For all the model components, we also report the improvement in the $\chi^{2}$ values that indicates how significant these components are in the spectral fit. A {\it constant} component was added to take into account the relative normalization of \suzaku{} and \nustar{} instruments. In XSPEC the model reads as: constant$\times$tbabs$\times$(po+diskbb+zgauss+pexrav). Below we discuss the different epoch of observations using different telescopes. 

\subsubsection{\xmm{}}

For obs1 and obs2, the two \xmm{} observations, fitting the $2 - 5\kev$ energy band with an absorbed power-law model and extrapolating to the rest of the X-ray band revealed a prominent soft excess below $2\kev$ (See APPENDIX \ref{appendix_softexcess} for details). The addition of {\it diskbb} component improved the fit statistics considerably ($\dc \sim 20$) for both these observations. The best-fit value of inner disk temperature is consistent with a best-fit value of $0.23^{+0.05}_{-0.05} \kev$ and $0.20^{+0.03}_{-0.04} \kev$ for obs1 and obs2 respectively. Next, we added a {\it Gaussian} component to the best-fit model. We could not constrain the line width $\sigma$ for the FeK line emission. We obtained an upper limit of $0.10\kev$ and $0.06\kev$ for obs1 and obs2 respectively. The best-fit power-law photon index values were consistent between observations.  All best-fit parameter values along with their fit-statistics are quoted in Table~\ref{Table:pheno}.

\subsubsection{\suzaku{}}

Obs3 and obs4, the two \suzaku{} observations, are fitted with similar a set of models. We fitted the $2-10\kev$ energy band with absorbed power-law model and did not find any excess emission in the source spectra (See APPENDIX \ref{appendix_softexcess} for details). To determine the upper limit on soft excess flux, we added a {\it diskbb} component to the absorbed power-law model. We could not constrain the inner disk temperature and for better comparison, fixed it to $0.20\kev$, the best-fit value we got from the spectral fit of obs2. Obs2 was preferred for its longer exposure. We found an excess emission around $6\kev$ for both obs3 and obs4 and added a {\it Gaussian} component to the set of models. We found a poor constraint on the line width $\sigma$ for obs4 ($0.09^{+0.09}_{-0.07}\kev$) and only an upper limit of $<0.12\kev$ for obs3. We investigated the hard X-ray band above $10\kev$ for Compton hump and modelled the data with {\it pexrav}. We found only an upper limit of the reflection fraction value (R) for obs3 and for obs4 this value was poorly constrained ($0.35^{+0.35}_{-0.30}$). The improvement in statistics was not significant (See Table~\ref{Table:pheno}).

\subsubsection{\nustar{} and \swift{}}

We used four (obs5, obs6, obs7, obs8) simultaneous \swift{} and \nustar{} observations. We followed the same spectral fitting procedure mentioned above and started with the absorbed power-law model. We did not find any soft excess emission and addition of {\it diskbb} does not improve the fit-statistics for any of the observations. Interestingly, for obs7 and obs8, with longer exposures, we could constrain the high energy cutoff of the power-law component (See APPENDIX \ref{appendix_softexcess} for details)). We used cut-off power-law model and found a lower limit of the electron temperature to be $>95\kev$ and $>79\kev$. The Fe emission line was modelled with a {\it Gaussian} component and we found upper limits of $\sigma<0.45\kev$ and $\sigma<0.44\kev$ on the emission line width for obs5 and obs6 respectively. We were able to constrain the line width for obs7 and obs8 with $\sigma=0.24^{+0.33}_{-0.16}\kev$ and $0.19^{+0.14}_{-0.10}\kev$ respectively. We did not find any positive residual above $10\kev$ in any of the \nustar{} observations. As a result, addition of {\it pexrav} component did not improve the fit statistics (See Table~\ref{Table:pheno}).

\subsubsection{Summary of results}

With our phenomenological modelling of the source spectra we found the presence of soft excess in the \xmm{} observations and a relatively weak and narrow ($\sigma < 0.4\kev$) Fe emission line. We did not find the presence of any obscuration or Compton hump above $10 \kev$ in the source spectra. We use physically motivated models next to investigate these spectral features in detail.


\begin{table*}
\centering

 \caption{The best fit parameters of the baseline phenomenological models for the observations of Mrk~590. \label{Table:pheno}}
{\renewcommand{\arraystretch}{1.2}
\setlength{\tabcolsep}{1.7pt}
\begin{tabular}{cccccccccc} \hline\hline
	
Models 		& Parameter 				&obs1 		&obs2	                 &obs3 		   &obs4 		&obs5 		 &obs6  &obs7   &obs8\\ \hline 

Gal. abs.  	& $\nh \,(\times 10^{20}\, \cmsqi)$ 	& $ 2.77$ (f)     	&$ 2.77$ (f)             & $ 2.77$ (f)     	   & $ 2.77$ (f)     	& $ 2.77$ (f)     	 & $ 2.77$ (f) & $ 2.77$ (f) &$2.77$ (f) \\

 powerlaw 	& $\Gamma$         			& $1.79^{+0.07}_{-0.03}$ &$1.76^{+0.04}_{-0.04}$ & $1.70^{+0.01}_{-0.01}$ & $1.70^{+0.01}_{-0.01}$ & $1.64^{+0.10}_{-0.09}$  & $1.66^{+0.10}_{-0.09}$ & $1.64^{+0.05}_{-0.04}$ & $1.61^{+0.05}_{-0.05}$ \\
 
        	& norm ($10^{-3}$) 			& $1.19^{+0.09}_{-0.12}$ &$1.56^{+0.06}_{-0.06}$& $1.78^{+0.02}_{-0.02}$& $1.59^{+0.02}_{-0.02}$ & $0.62^{+0.10}_{-0.08}$  & $0.65^{+0.10}_{-0.09}$ & $2.86^{+0.16}_{-0.16}$ & $1.38^{+0.10}_{-0.09}$ \\
        	& $E_{Cutoff}$ ($\kev$)     & $--$    &$--$       & $--$      & $--$          & $--$        & $--$      & $>95$       & $>79$ \\

diskbb  	& $T_{in}$ ($\kev$)   			&$0.23^{+0.05}_{-0.05}$ &$0.20^{+0.03}_{-0.04}$ & $0.20$(f)             & $0.20$(f)              & $0.20$(f)           & $0.20$(f) & $0.20$(f) & $0.20$(f)  \\
          	& norm           			&$22.4^{+7.1}_{-7.6}$   &$28.5^{+3.7}_{-10.7}$  & $7.1^{+5.5}_{-5.6}$   & $9.9^{+5.2}_{-6.6}$    & $<11.9$             & $<19.5$ & $<5.1$ & $<25.8$   \\
		&$\rm ^A$$\rm \dc/dof$			&$21/2$		 &$26/2$                 &$4/1$		   &$9/1$		     &$1/1$		 &$1/1$	         &$1/1$                 &$3/1$\\

Gaussian	&E($\kev$)        			& $6.39^{+0.05}_{-0.06}$ &$6.41^{+0.03}_{-0.03}$ & $6.42^{+0.03}_{-0.03}$ & $6.42^{+0.05}_{-0.04}$ & $6.48^{+0.21}_{-0.16}$ & $6.48^{+0.22}_{-0.16}$ & $6.35^{+0.17}_{-0.08}$ & $6.32^{+0.08}_{-0.08}$ \\
		&$\sigma$($\kev$)        		& $<0.10$                &$<0.06$               & $<0.12$		   & $0.09^{+0.09}_{-0.07}$ & $<0.45$		      & $<0.44$ &$0.24^{+0.33}_{-0.16}$ &$0.19^{+0.14}_{-0.10}$ \\
		&norm ($10^{-5}$) 			& $1.05^{+0.51}_{-0.46}$ &$0.78^{+0.23}_{-0.24}$ & $1.23^{+0.26}_{-0.25}$ & $1.02^{+0.37}_{-0.30}$ & $0.73^{+0.46}_{-0.39}$ & $0.72^{+0.45}_{-0.39}$ & $2.61^{+1.35}_{-0.71}$ & $1.68^{+0.49}_{-0.43}$ \\

		&$\rm ^A$$\rm \dc/dof$			&$17/3$		 &$33/3$                 &$99/3$		   &$53/3$		    &$14/3$		 &$15/3$	&$75/3$       &$54/3$ 	\\

Pexrav $^{B}$           & R         & $<0.57$            &$0.31$ (t)            & $<0.24$              &$0.35^{+0.35}_{-0.30}$ & $<0.44$          &$<0.53$    &$--$      &$--$        \\
			& Incl(degree)             		& $12$(t)            &$12$ (t)              & $12$(*)              & $12$(t) 		        & $12$(t)        & $12$(t)     & $--$           & $--$ \\
			&$\rm ^A$ $\rm \dc/dof$ & $4/2$                 & $7/2$               & $2/2$                 & $4/2$ 		        & $3/2$      	  & $4/2$        & $--$          & $--$  \\

Gaussian  	&  EqW (eV)  				 & 174                    & 123                 & 150                     & 132 	    &  227            & 169      & 182      & 229      \\\hline 

$\cd$     &                  				& $ 104/122$            &$ 201/161 $          & $ 907/831 $            & $ 555/549 $            & $ 193/208 $    & $ 304/313 $   & $ 878/837 $  & $ 572/613 $ \\\hline
\end{tabular} }\\ 

{$\rm ^A$ The $\dc$ improvement in statistics upon addition of the corresponding discrete component.}\\
{$\rm ^B$ The model {\it pexrav} was used only for \suzaku{} and \nustar{} observation as it had broad band spectra necessary for constraining the parameters. The values quoted for the \xmm{} observations are from the simultaneous fit of all the data sets.\\
 R represents the reflection component only.\\
 The temperature at inner disk radius $T_{in}$ (keV) for \suzaku{} and \swift{} $+$ \nustar{} observations, when made free, was taking very low values and hence was fixed at $0.2\kev$. \\
 (*) indicates parameters are not constrained}\\

\end{table*}

\subsection{The physical models}\label{subsec:phys}
 
 \subsubsection{Ionized disk reflection}
 
We used the {\it relxill} model, version 1.4.0, \citep{2014ApJ...782...76G,2014MNRAS.444L.100D} in our spectral fitting. This model assumes the origin of soft excess to be relativistic reflection from an ionized accretion disk or simply ionized reflection. We added the MyTorus model \citep{2010MNRAS.401..411Y, 2011MNRAS.412..277Y} to take into account the distant neutral reflection from outer part of the disk or torus. In {\sc XSPEC}, our model reads as constant$\times$tbabs$\times$(relxill+MyTorus). The {\it relxill} model describes the soft X-ray excess emission, the X-ray continuum and the broad Fe K$_{\alpha}$ emission line. The distant neutral reflection on the other hand is modelled with the two {\it MyTorus} model components, first {\it MyTorusL}, which describes the iron Fe K$_{\alpha}$ and K$_{\beta}$ lines and second the {\it MyTorusS}, which models the scattered emission due to the reﬂection of primary power-law emission from the torus. The best fit parameters of model components for all observations are quoted in Table~\ref{Table:relxill}, along with their chi-squared fit statistic.

The {\it relxill} model assumes a lamp-post geometry of the corona where part of the hard X-ray continuum enters the accretion disk, ionizes it and emits fluorescence lines. These emission lines then get blurred and distorted due to the extreme gravity around the central super massive black hole (SMBH) and along with scattered emission from ionized accretion disk, produces the soft excess emission and a broad Fe emission line around $6\kev$. The transition between this relativistic and Newtonian geometry is characterized by a breaking radius $r_{br}$. We fix the emissivity index of reﬂection from the disk outside this $r_{br}$ (q2) at 3, as expected for a point source under Newtonian geometry at a large distance from the source. Whereas, the emissivity inside the radius $r_{br}$, q1, was allowed to vary as this region falls under the relativistic high-gravity regime and previous studies \citep{2001MNRAS.321..605D,2003MNRAS.344L..22M,2011MNRAS.414.1269W} suggest a very steeply falling proﬁle in the inner parts of the disk. In our spectral fits with phenomenological models we did not detect any broad Fe emission line which would indicate an rotating black hole. To confirm the non-detection, we tested two extreme scenarios. First, we fixed the black hole spin parameter to maximum value of 0.998 and the inner radius ($r_{\rm in}$) to $1.24 r_{\rm g}$, lowest value allowed in the model, and in the second scenario, we fixed the spin to zero and inner radius to $6 r_{\rm g}$. We found that the fit-statistics is insensitive to both rotating and non-rotating scenarios for all our observations. Since the data are insensitive to the spin of the black hole, we continued with the non-rotating black hole scenario in all the spectral fits in our work. Accordingly, we fixed $r_{\rm br}$ to a larger value of 10$r_{\rm g}$ and,  q1, the emissivity index inside $r_{\rm br}$ to 3. 

\begin{figure*}
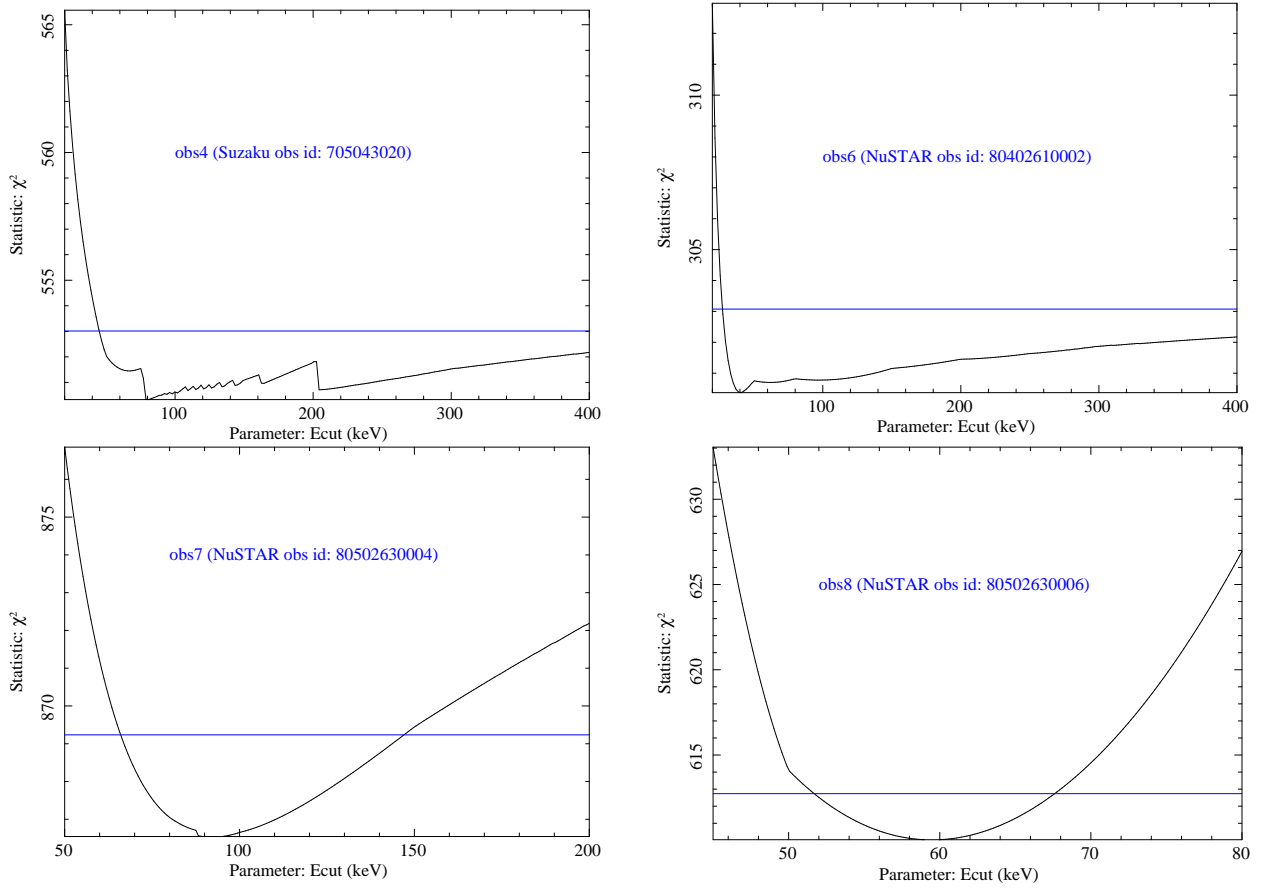

  \centering 
  
\hbox{
\includegraphics[angle=-90,width=8.5cm]{Ecut_contour_obs4.eps}
\includegraphics[angle=-90,width=8.5cm]{Ecut_contour_obs6.eps}}
\hbox{
\includegraphics[angle=-90,width=8.5cm]{Ecut_contour_obs7.eps}
\includegraphics[angle=-90,width=8.5cm]{Ecut_contour_obs8.eps}}

\caption{The confidence intervals plot of the high energy cut-off parameter $\rm E_{cut}$ of four observations, obs4, obs6, obs7 and obs8. Top panel shows the confidence contour plot of obs4 and obs6 among others that could not be constrained due to low-quality data above $20\kev$. We found similar result for obs3 and obs5. Bottom panel shows the same contour plot for obs7 and obs8 which are well constrained. } \label{fig:confidence_contour_ecut}

\end{figure*}

The availability of hard X-ray data beyond $10\kev$ for obs3 to obs8 involving \suzaku{}, \swift{} and \nustar{}, provided us an opportunity to measure the high-energy cut-off in this changing-look AGN. We made the parameter $\rm E_{cut}$ free in all our fits. However, we could not constrain the value of the $\rm E_{cut}$ for obs3 to obs6. Interestingly, we get a well-constrained high energy cut-off of $92^{+65}_{-25}\kev$  and $60^{+10}_{-08}\kev$ for obs7 and obs8, respectively. To robustly identify the errors on the energy cut-off, we carried out some statistical tests. Using the {\sc steppar} command in {\sc XSPEC}, we determined the confidence intervals for the parameter. In Fig.~\ref{fig:confidence_contour_ecut}, (top panel) we show the example of obs4 and obs6 where the $\rm E_{cut}$ could not be constrained. The bottom panel shows the confidence intervals of obs7 and obs8, where $\rm E_{cut}$ is well constrained. Obs1 and obs2 involving \xmm{} do not cover above $10 \kev$ energy band. We tested with different $\rm E_{cut}$ values (60, 90, and $300\kev$) and noted that the fit is insensitive to the parameter and hence fixed the parameter to $300\kev$ for these two observations.

The iron abundance of the material in the accretion disk is represented by the parameter $A_{Fe}$. We first tied this parameter between obs1 and obs2 and only got an upper limit ($<0.78$); hence, we fixed it to solar values. For other observations, when made free, we found it to be pegged at highest allowed value of 5. Hence, we fixed it to solar abundance value for obs3 to obs8 and it did not affect the fit statistics($\dc = -3$). 

The inclination angle (in degrees) found for \xmm{} observations are consistent with a best-fit value of $45^{+5}_{-7}$ degree. The inclination angle is not supposed to change during the human time scale, and we froze this parameter to 45 degrees for all the other observations.

The ionization of the accretion disk is characterized by the parameter $\log \xi$, which varied significantly between observations. Due to better data quality, we were able to constrain $\log\xi$ to $0.52^{+0.77}_{-0.30}\xiunit$ for obs2 and got an upper limit of 2.01 for obs1. For the rest of the observations we got a highly ionized accretion disk with the best-fit parameter value ranging between $2.72^{+0.18}_{-0.68}\xiunit$ in obs4 and $3.30^{+0.91}_{-1.45}\xiunit$ in obs8 but remained consistent within errors. 

The reflection fraction (R) determines the ratio of photons emitted towards the disk compared to escaping to infinity. We obtained relatively higher values ($\sim 0.46$) for obs1 and obs2. However, we note that the best-fit values of the reflection fraction are poorly constrained and are within the $3\sigma$ limit.

For the model {\it MyTorus} we fixed the inclination angle to 45 degrees and tied the column density of both {\it MyTorusL} and {\it MyTorusS} in all our observations. Due to the lack of data beyond $10\kev$, in \xmm{} observations, {\it MyTorus} normalization was tied between {\it MyTorusL} and {\it MyTorusS}. We were unable to constrain the column density and found it to be pegged at 10$\times10^{24}\cmsqi$ for all observations. Hence, we fixed this value to 10 $\times10^{24}\cmsqi$ for all our observations and made flux normalization the only free parameter during our spectral fitting. 

The optical/UV data from simultaneous \xmm{} OM and \swift{} UVOT instruments were modelled with {\it diskbb}.  To avoid the effects of host galaxy and starburst contribution we did not consider the V and B band in our spectral fits, which are more likely to be affected by these phenomenon. Further, the optical/UV flux may be contaminated due to emission from the BLR/NLR region. Unfortunately, the contamination cannot be quantified accurately from the \xmm{} OM data. Using Hubble Space Telescope (HST) data, \citet{2018ApJ...866..123M} measured the BLR continuum flux to be $\sim7-10$\% of the UV continuum flux. Following this measurement, we corrected the count rates in the UV and derived the intrinsic count rates of the source. We wrote these count rates in an OGIP compliant spectral file generated using {\it om2pha} and {\it uvot2pha} tasks for the \xmm{} and \swift{}, respectively. We also introduced a typical 5\% systematic uncertainty \citep{2013ApJ...777....2L,2018MNRAS.479.2464G} to the optical-UV data sets for each epoch to account for the intrinsic galactic extinction and the host galaxy contribution. Here, we froze the {\it relxill} and {\it MyTorus} model parameters to their best-fit values obtained from the $0.3-50\kev$ X-ray spectral fitting. We included a {\it REDDEN} component to account for the inter-stellar extinction. The {\it diskbb} model describes the optical/UV band well and provides a satisfactory fit for all the observations (See APPENDIX \ref{appendix_relxill} for details).


\begin{table*}

\centering
  \caption{Best fit parameters for observations of Mrk~590 with the first set of physical models. In XSPEC, the models read as {\it(constant $\times$ tbabs$\times$(relxill + MYTorus))}. \label{Table:relxill}}

{\renewcommand{\arraystretch}{1.2}
\setlength{\tabcolsep}{1.8pt}
  \begin{tabular}{cccccccccc} \hline
Component  & parameter                	     & obs1           &obs2                   & obs3 		& obs4 		  & obs5		& obs6                     &obs7      &obs8\\\hline

Gal. abs.  & $\nh (10^{20} cm^{-2})$      & $ 2.77$ (f)             &$ 2.77$ (f)            & $ 2.77$ (f)           & $ 2.77$ (f)     	  & $ 2.77$ (f)        & $ 2.77$ (f) & $ 2.77$ (f) & $ 2.77$ (f) \\ 

relxill     &  $A_{Fe}$               & $1$ (f)  	       &  $1$ (f)              & $1$(f)                & $1$ (f)   	          & $1$(f)                        & $1$(f)       & $1$(f)      & $1$(f) \\
 
 	        &  $\log\xi (\xiunit)$     & $<2.01$                 &$0.52^{+0.77}_{-0.30}$ & $3.19^{+0.45}_{-0.84}$& $2.72^{+0.18}_{-0.68}$&$3.19^{+0.49}_{-0.47}$& $3.30^{+0.91}_{-1.45}$ & $3.11^{+0.46}_{-0.80}$ & $3.30^{+0.55}_{-0.99}$ \\ 
            & $ \Gamma $               & $1.88^{+0.02}_{-0.08}$  &$1.81^{+0.02}_{-0.05}$ & $1.68^{+0.02}_{-0.01}$&$1.65^{+0.06}_{-0.01}$ &$1.62^{+0.10}_{-0.06}$ &$1.60^{+0.04}_{-0.03}$ &$1.60^{+0.06}_{-0.06}$ &$1.58^{+0.02}_{-0.03}$ \\
            & $\rm E_{cut}(\kev)$      & $300$ (f)               &$300$(f) 	        & $>32$              &$>73$               & $>28$             & $>65 $ & $92^{+55}_{-25}$ & $60^{+10}_{-8}$ \\

            & $\rm E_{cut}(\kev)$      & $300$ (f)               &$300$(f) 	        & $55$(f)              &$90$ (f)              & $52$ (f)             & $ 70$ (f) & $92^{+55}_{-25}$ & $60^{+10}_{-8}$ \\

           &  $n_{rel}(10^{-5})^a$    & $2.44^{+0.38}_{-0.12}$  &$3.71^{+0.55}_{-0.18}$ &$5.32^{+0.28}_{-0.75}$ &$5.28^{+0.17}_{-0.17}$ &$2.17^{+0.45}_{-0.43}$ &$7.27^{+0.97}_{-0.81}$ &$6.20^{+1.32}_{-1.20}$ &$3.11^{+1.11}_{-1.81}$ \\
	   
           	&   $ q1$                  & $3$(f)            &$3$(f)                 & $3$(f)                & $3$(f)                &$3$(f)                &$3$(f)         &$3$(f)   &$3$(f) \\
 
		    &   $ a$                   & $0$(f)    	       &$0$(f)                 & $0$(f)                & $0$(f)                & $0$(f)               &$0$(f)         &$0$(f)   &$0$(f)\\
		
           	&   $R_{\rm frac} $    & $0.47^{+0.34}_{-0.40}$  &$0.46^{+0.16}_{-0.26}$ & $0.04^{+0.06}_{-0.02}$&$0.08^{+0.05}_{-0.04}$&$0.14^{+0.28}_{-0.10}$ & $<0.14$ &$0.24^{+0.27}_{-0.13}$ &$0.24^{+0.32}_{-0.12}$ \\

           	&   $ R_{in}(r_{g})$       & $6$(f)                  &$6$(f)                 & $6$(f)                & $6$(f)                & $6$(f)               & $6$(f)       & $6$(f)     & $6$(f) \\

		    &   $ R_{br}(r_{g})$       & $10$(f)                 &$10$(f)                & $10$(f)               & $10$(f)               & $10$(f)              & $10$(f)       & $10$(f)  & $10$(f) \\

           	&   $ R_{out}(r_{g})$      & $400$ (f)      	     &$400$(f)               & $400$(f)              & $400$(f)              & $400$(f)             & $400$(f)   & $400$(f)     & $400$(f) \\

           	&   $i(\rm degree) $       & $45$(f)	             &$45^{+5}_{-7}$         & $45$(f)                 & $45$(f)                 & $45$(f)              & $45$(f)   & $45$(f)  & $45$(f) \\\hline

{\it MYTorusL}  &   $i(\rm degree) $       & $45$(f)                &$ 45$(f)                & $ 45$(f)              & $45$(f)              & $45$(f)               & $45$(f)  & $45$(f)  & $45$(f) \\

		        &  norm ($10^{-3}$)        & $6.88^{+2.71}_{-2.69}$ &$4.07^{+1.27}_{-1.26}$  &$4.09^{+0.65}_{-0.70}$ &$2.80^{+0.70}_{-0.76}$ &$1.46^{+1.02}_{-0.82}$ & $4.58^{+2.03}_{-1.82}$ & $4.69^{+1.70}_{-1.52}$ & $3.01^{+1.33}_{-1.05}$ \\

{\it MYTorusS } &  NH($10^{24}\rm cm^{-2}$)& $10$(f)                &$10$(f)                 & $10$(f)               &$10$(f)               &$10$(f)                & $10$(t)   & $10$(t)  & $10$(t) \\

		        &  norm ($10^{-3}$)        & $6.88$(f)               &$4.07$(f)               &$<0.50$               &$<0.62$                &$<0.55$               &$<0.75$     &$<0.71$  &$<0.50$   \\\hline
		
	   	        & $\cd $                   & $108/127$      	      &$206/163$               &$899/833$             &$551/550$              &$191/209$           & $302/314$  & $867/836$   & $578/612$  \\\hline 
With OM data &&&& & &&&&\\\hline 

diskbb          & $T_{in}$ (eV)           &$1.25^{+0.04}_{-0.04}$   &$0.94^{+0.02}_{-0.02}$ & $--$                    & $--$            &$1.08^{+0.04}_{-0.04}$    & $1.75^{+0.09}_{-0.07}$   & $1.62^{+0.08}_{-0.07}$  & $1.41^{+0.06}_{-0.05}$ \\
                & norm ($\times 10^{12}$) &$0.30^{+0.06}_{-0.05}$   &$1.59^{+0.18}_{-0.16}$ & $--$                    & $--$            &$0.60^{+0.15}_{-0.12}$    & $0.13^{+0.04}_{-0.03}$   & $0.15^{+0.04}_{-0.03}$  & $0.20^{+0.05}_{-0.04}$ \\\hline

                & $\cd $                   & $164/134$      	      &$222/171$               &$--$             &$--$              &$230/214$           & $344/317$  & $910/845$   & $580/621$  \\\hline 
\end{tabular}\\
}
Notes: Spectral fitting of all observations include simultaneous optical-UV data except obs3 and obs4.\\ (f) indicates a frozen parameter. (t) indicates a tied parameter between observations. \\(a) $n_{rel}$ represent normalization for the model {\it relxill}.\\

\end{table*}


 \subsubsection{Warm Comptonization}

In our work, we used {\it optxagnf} as the warm Comptonization model to describe the soft excess emission. {\it Optxagnf}~\citep{2012mnras.420.1848d} is an intrinsic thermal Comptonization model which describe (a) the optical/UV spectra of AGN as multi-colour black body emission from colour temperature corrected disk, (b) the soft X-ray excess emission as thermal Comptonisation of disk seed photons from a optically thick, low temperature plasma, and (c) power-law continuum as thermal Comptonisation of disk photons from an optically thin, hot (fixed at $100 \kev$) plasma. All three components are powered by the gravitational energy released due to accretion. $\rm r_{corona}$ determines the inner radius below which the gravitational energy can not completely thermalise and is distributed among the soft X-ray excess and the power-law components. This ratio is determined by the $\rm f_{pl}$ parameter. The electron temperature ($kT_{e}$) and optical depth ($\tau$) represents the warm corona responsible for the soft excess emission. The model flux is determined by four parameters, the black hole mass ($\rm M_{BH}$), the Eddington ratio ($\lambdaedd$), the co-moving distance (D in $\mpc$) and the dimensionless black hole spin ($\rm a$). Hence the model normalisation is fixed at unity during spectral fitting. Similar to {\it relxill}, we included the two {\it MyTorus} model components, {\it MyTorusL} and {\it MyTorusS}, to our set of models to account for the neutral reflection of hard X-ray continuum from the outer part of the disk. 
 
We fixed the black hole mass of Mrk~590 to $4.75\times 10^7\msol{}$, determined using the reverberation mapping  \citep{2004ApJ...613..682P}, and the cosmological distance to $112.88\mpc$~\citep{2000ApJ...529..786M}. {\it Optxagnf} model needs optical-UV data to constrain the multi colour black body emission from disk. Hence, the model resulted a poor constraint in parameters when we fitted only the X-ray band with this set of models. To get a better constrain we included the simultaneous optical-UV data from \xmm{} and \swift{} telescopes for all observations except for obs3 and obs4 where we did not have simultaneous data. 
 
Similar to previous set of physical models, the fit-statistic was insensitive to the black hole spin for all observations. Hence we fixed the spin parameter to zero and allowed the Eddington ratio ($\rm L/L_{Edd}$), the optical depth ($\tau$), the electron temperature ($\rm kT_{e}$), the photon index ($\Gamma$) and the $\rm f_{pl}$ parameter to vary freely. For the {\it MyTorus} model components we fixed the inclination angle to 45 degrees and allowed the {\it MyTorusL} and {\it MyTorusS} normalization parameters to vary freely except for obs1 and obs2 where we tied them together. Similar to the case of {\it relxill} model, we found the {\it MyTorus} column density to be pegged at 10$\times10^{24}\cmsqi$ for all observations and hence, fixed this value to 10$\times10^{24}\cmsqi$ for all the spectral analysis. The best-fit parameters are quoted in Table~\ref{Table:optxagnf}.
 
\begin{figure*}
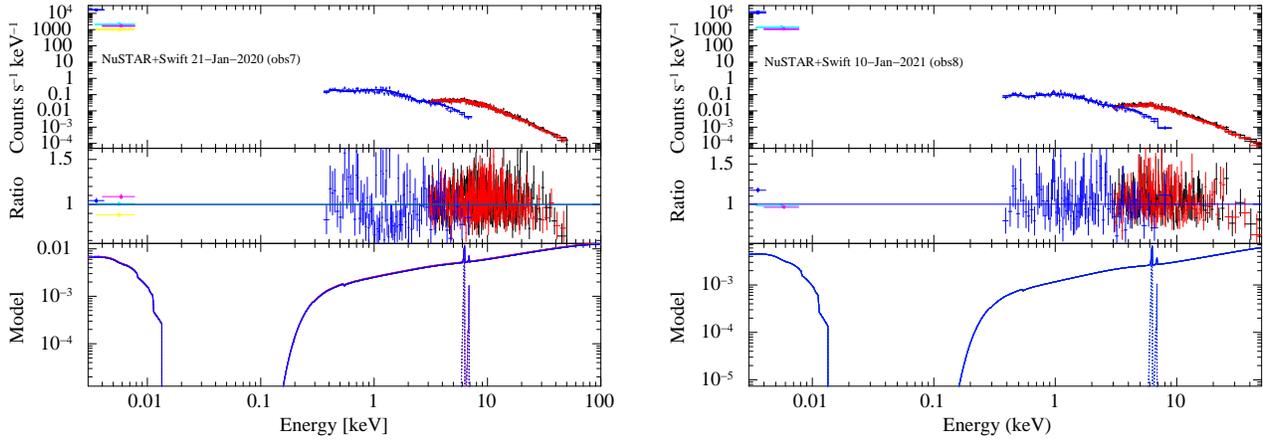

  \centering 

\hbox{
\includegraphics[width=6.0cm,angle=-90]{best_fit_abs_optxagnf_mytorus_withoutspin_uvot_obs7.eps}
\includegraphics[width=6.0cm,angle=-90]{best_fit_abs_optxagnf_withoutspin_mytorus_uvot_obs8.eps}}

\caption{The $0.001-50.0\kev$ simultaneous \swift{} and \nustar{} spectra of Mrk~590 fitted with an absorbed {\it optxagnf} and {\it MyTorus} model. The broadband data, the residuals and the theoretical model shown for obs7 and obs8. We see the set of models fail to describe the high energy cut-off of primary continuum observed above $20\kev$. We fitted each data sets separately. The X-axis represents observed frame energy.} \label{fig:optxagnf_fit}

\end{figure*}

 The {\it optxagnf} model produced comparable fit statistics of the broadband source spectra for all the observations except for obs7 and obs8. For obs7 and obs8, we find this set of models provided a poor description of the observed high-energy cut-off above $20\kev$. Fig.~\ref{fig:optxagnf_fit} shows the residuals and the theoretical model for these two observations. The {\it optxagnf} model can describe the UV bump and hence a separate {\it diskbb} is not required in principle. However, given that we get poor fit using the {\it optxagnf} alone for obs1, obs2 and obs5, we added a separate {\it diskbb} component which improved the fit statistics by $\dc \sim 120-130$. Clearly the optical-UV data requires this additional component. We notice that the optical-UV flux measured during obs1, obs2 and obs5 to be relatively lower compared to obs6, obs7 and obs8. However, we were unable to constrain the optical depth ($\tau$) and the coronal radius ($\rm r_{corona}$) for most of the observations. We found a sub-Eddington accretion rate ($1-3\%$) in all observations. The variability in power-law photon index ($\Gamma$) between observations were not statistically significant. We found very high values of $\rm f_{pl}$ in all observations except for obs6. Interestingly, for obs6 we found the electron temperature ($\rm kT_{e}$) to be very low compared to other observations. The {\it MyTorusL} flux normalization values were consistent between observations. We could not constrain the {\it MyTorusS} flux normalization and only got an upper limit on the flux.

 \subsubsection{Summary}
 
Our spectral analysis shows that both set of physical models provide a satisfactory fit to the source spectra and we can not distinguish them on fit statistics alone for most of the observations. For obs7 and obs8, where we found the presence of a high energy cut-off, the {\it relxill} plus {\it MyTorus} model provided a better fit-statistics ($\dc=50$). 


\begin{table*}

\centering
  \caption{Best fit parameters for observations of MRK~590 with the second set of physical models. In XSPEC, the models read as {\it(constant $\times$ tbabs$\times$(diskbb + optxagnf+ MYTorus))}.\label{Table:optxagnf}}
{\renewcommand{\arraystretch}{1.2}
\setlength{\tabcolsep}{1.5pt}
  \begin{tabular}{cccccccccc} \hline

Component  & parameter                & obs1     	       & obs2                  & obs3		         & obs4		& obs5 		& obs6     & obs7     & obs8  \\\hline

Gal. abs.  & $\nh (10^{20} cm^{-2})$ & $ 2.77$ (f)           &$ 2.77$ (f)            & $ 2.77$ (f)             & $ 2.77$ (f)     	& $ 2.77$ (f)           & $ 2.77$ (f)   & $ 2.77$ (f)  & $ 2.77$ (f)\\

diskbb     & $T_{in}$ (eV)         &$1.22^{+0.12}_{-0.20}$   &$0.78^{+0.03}_{-0.05}$ & $--$                    & $--$              &$0.25^{+0.05}_{-0.02}$    & $--$   & $--$  & $--$ \\
           & norm ($\times 10^{12}$) &$0.25^{+0.22}_{-0.08}$   &$2.92^{+1.25}_{-0.56}$ & $--$                    & $--$              &$880^{+112}_{-595}$             & $--$   & $--$  & $--$ \\
           &$\rm ^A$ $\rm \dc/dof$ &$91/2$		             &$311/2$                & $--$		               & $--$		       &$140/2$		              & $--$   & $--$  & $--$ \\

optxagnf   & $ \rm M_{BH}^a$       &$4.75$(f)                 &$4.75$(f)               & $4.75$(f)                &$4.75$(f)               &$4.75$(f)               &$4.75$(f)  &$4.75$(f)  &$4.75$(f) \\
           & $d{\rm~(Mpc}) $       &$113$(f)                 &$113$(f)               & $113$(f)                &$113$(f)               &$113$(f)               &$113$(f)  &$113$(f)  &$113$(f) \\
           & $(\frac{L}{L_{E}})$   &$0.006^{+0.003}_{-0.001}$&$0.009^{+0.001}_{-0.002}$& $0.22^{+0.01}_{-0.01}$&$1.20^{+0.38}_{-1.05}$ &$0.005^{+0.003}_{-0.001}$ &$0.030^{+0.002}_{-0.002}$  &$0.020^{+0.001}_{-0.001}$  &$0.012^{+0.001}_{-0.001}$ \\ 
           & $ kT_{e} (\kev)$      &$0.18^{+0.14}_{-0.08}$   &$0.17^{+0.05}_{-0.04}$ & $0.05^{+0.27}_{-0.01}$  &$0.05$(t)              &$>0.48$                &$0.03^{+0.01}_{-0.01}$ &$>0.57$  &$>0.55$  \\ 
           & $ \tau $              &$>16$                    &$>23$                  & $>9$                    &$99$(t)                &$>6$                   &$>71$   &$>40$   &$>31$   \\
           & $ r_{\rm cor}(r_{g})$ &$>63$                    &$>93$                  & $9.8^{+0.1}_{-0.1}$     &$>7.2$                 &$55^{+10}_{-6}$      &$>89$   &$78.8^{+0.1}_{-0.1}$   &$64.7^{+0.9}_{-1.3}$ \\
           & $ a $                 &$0$(f)                   &$0$(f)                 & $0$(f)                  &$ 0$(f)                &$ 0$(f)                &$0$(f)  &$0$(f)  &$0$(f) \\
           & $ f_{pl}$             &$0.97^{+0.02}_{-0.11}$   &$0.97^{+0.01}_{-0.03}$ & $0.62^{+0.37}_{-0.01}$  &$>0.52$                &$0.98^{+0.01}_{-0.03}$ &$0.50^{+0.06}_{-0.04}$ &$0.99^{+0.01}_{-0.01}$ &$0.98^{+0.01}_{-0.01}$ \\
           & $ \Gamma $            &$1.78^{+0.11}_{-0.16}$   &$1.74^{+0.05}_{-0.05}$ &$1.71^{+0.01}_{-0.01}$   &$1.68^{+0.01}_{-0.01}$ &$1.62^{+0.06}_{-0.08}$ &$1.66^{+0.03}_{-0.02}$  &$1.65^{+0.02}_{-0.02}$ &$1.61^{+0.01}_{-0.01}$ \\\hline

{\it MYTorusL}  & $i(\rm degree) $         & $45$ (f)                 &$ 45$(f)       & $ 45$(f)                & $ 45$(f)             & $45$(f)               & $45$(f)   & $45$(f)  & $45$(f)\\
                & norm ($10^{-3}$)  & $5.62^{+2.81}_{-2.29}$  &$4.59^{+1.21}_{-1.16}$ & $4.38^{+0.73}_{-0.72}$  &$3.31^{+0.79}_{-0.76}$&$2.09^{+1.05}_{-0.99}$ & $5.56^{+2.10}_{-2.05}$ & $7.67^{+1.26}_{-1.24}$ & $4.34^{+0.88}_{-0.04}$ \\

{\it MYTorusS } & NH($10^{24}\rm cm^{-2}$) & $10.0$(f)                &$10.0$(f)      & $10.0$(t)               &$10.0$(t)             &$10.0$(t)              & $10.0$(t)   & $10.0$(t)  & $10.0$(t)\\
                & norm ($10^{-3}$)         & $5.14$ (f)               &$3.00$(f)      & $<0.48$                 &$<1.09$               &$<0.76$         & $<0.52$     & $<0.05$    & $<0.06$ \\\hline
		
                & $\cd $                   & $160/124$                & $218/163$     &$898/831 $               & $546/550 $           &$ 189/209$             &$340/316 $   &$972/840 $  &$632/615$ \\\hline 
\end{tabular}
}\\ 
Notes: The quoted best-fit values for \suzaku{} observations (obs3 and obs4) are from the spectral fitting of X-ray band only.\\ (f) indicates a frozen parameter. (t) indicates a tied parameter between different observation.\\(*) indicates parameters are not constrained. \\(a):in units of $10^7\rm M\odot$;\\
\end{table*}


\section{Results}\label{sec:results}


\subsection{Soft excess variability}\label{subsec:soft}

\begin{figure*}

\centering{\includegraphics[width=18.5cm,height=22cm]{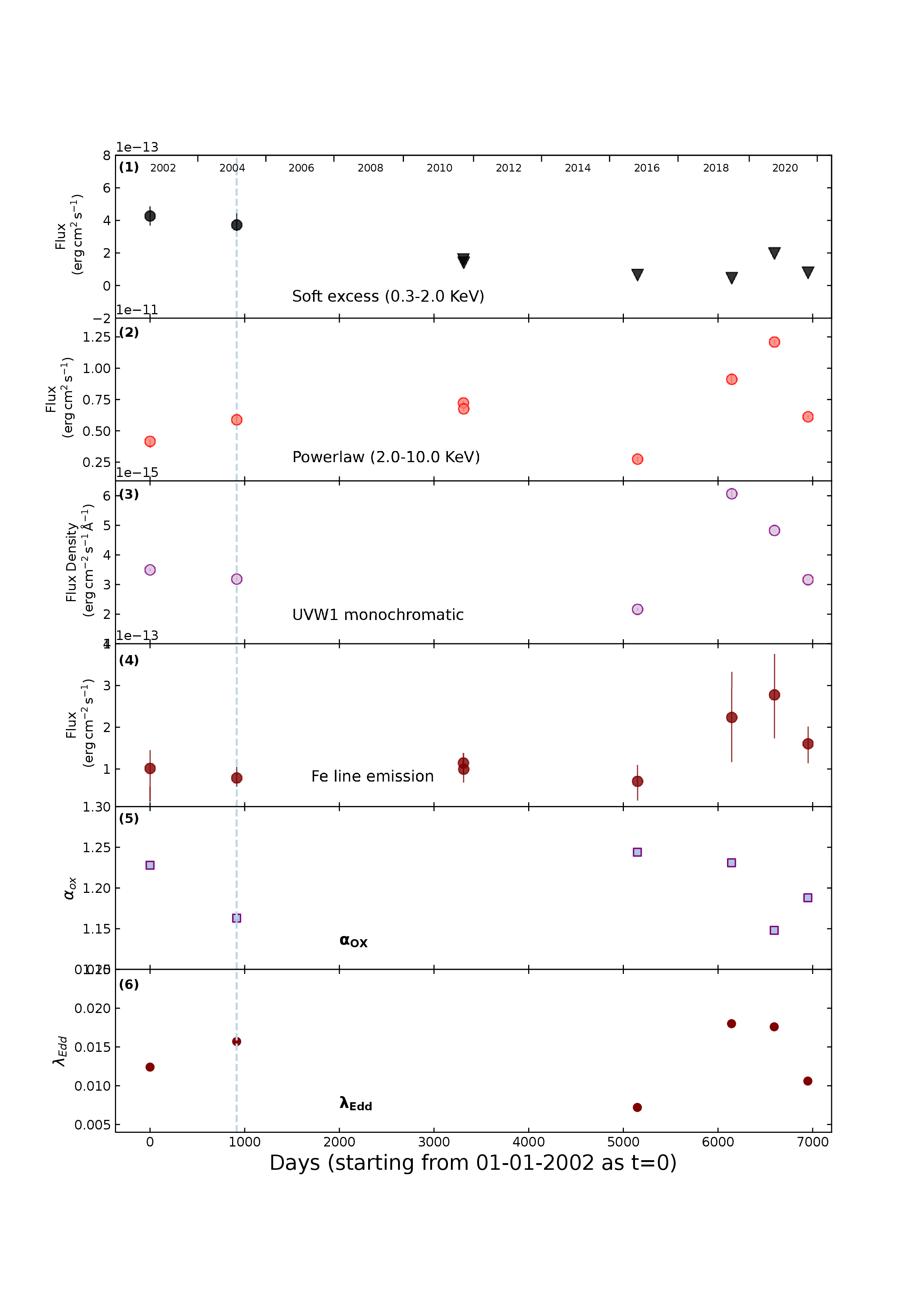}}

\caption{The X-ray and UV parameters of the central engine of the AGN Mrk~590, as observed by \xmm{}, \suzaku{}, \nustar{} and \swift{} (see Table~\ref{Table:flux} for details). The start date is 2002-01-01 corresponds to the first \xmm{} observation (obs1). The X-axis is in the units of days elapsed from the start date. From the top to the bottom are panels: (1) The soft X-ray excess flux in the $0.3-2\kev$ band (in units of $10^{-13} \funit$), (2) The power-law flux in the $2-10\kev$ band (in units of $10^{-13} \funit$), (3) The UV(UVW1) monochromatic flux density at $2500\mathrm \AA$ (in units of $10^{-15} \funit$), (4) The Fe line emission flux (in units of $10^{-13} \mathrm{erg\,cm^2\,s^{-1}\,\AA^{-1}}$), 
(5) The $\alpha_{OX}$ and (6) The Eddington ratio ($\lambdaedd$). The vertical line represents the epoch up to when soft excess is present.} \label{fig:variation_flux}

\end{figure*}

Our spectral analysis revealed a significant variability in soft excess flux between observations of Mrk~590. The soft excess was present ($3.7^{+1.0}_{-0.5}\times10^{-13}\funit$) in 2004 (obs2) but was vanished/undetected ($<1.6\times10^{-13}\funit$) in 2011 (obs3 and obs4), within a period of seven years. This excess emission never reappeared in any of the later observations till 2021. We calculated the $0.3-2.0\kev$ soft X-ray excess flux from the phenomenological best-fit and quoted these values in Table~\ref{Table:flux}. The soft excess flux ($F_{SE}$) in obs1 and obs2 are $F_{SE}= 4.3^{+0.6}_{-0.6}\times 10^{-13}\funit$ and $3.7^{+1.0}_{-0.5}\times 10^{-13}\funit$ respectively. We do not detect the soft excess in obs3 and obs4 and the corresponding upper limit are $F_{SE}< 1.6\times10^{-13}\funit$ and $<1.4\times10^{-13}\funit$ respectively. For obs5, obs6, obs7 and obs8 (\nustar{} plus \swift{} observations) the upper limit on the soft excess flux are $F_{SE}< 0.7\times10^{-13}$, $<0.6\times10^{-13}$, $<1.9\times10^{-13}$ and $<0.8\times10^{-13}\funit$ respectively. From Fig~\ref{fig:variation_flux} (top panel) we see that the soft excess flux drops by a factor of four within nine years, from 2002 to 2011.


\begin{table*}

\centering
  \caption{The fluxes of the different spectral components of Mrk~590 obtained from the observations used in our work. \label{Table:flux}}
  \setlength{\tabcolsep}{1.3pt}
  \begin{tabular}{cccccccccc} \hline\hline 

Spectral					                 &Flux			         & Flux		             & Flux		             & Flux		             & Flux		     & Flux       & Flux   & Flux \\
Component					                 &obs1			         & obs2 		         & obs3		             & obs4		             & obs5		     & obs6       & obs7   & obs8 \\
                                             &(JAN 2002)           &(JULY 2004)           &(JAN 2011)           &(JAN 2011)           &(FEB 2016)           &(OCT 2018)            &(JAN 2020) &(JAN 2021) \\\hline
Soft Excess  ($\times 10^{-13}$)		     &$4.27^{+0.63}_{-0.55}$ &$3.72^{+0.96}_{-0.48}$ &$<1.60$ &$<1.40$ &$<0.65$                &$<0.55$                 &$<1.97$      &$<0.79$\\
Power law$^1$    ($\times 10^{-12}$)    	 &$4.17^{+0.51}_{-0.19}$ &$5.89^{+0.28}_{-0.27}$ &$7.24^{+0.17}_{-0.16}$ &$6.76^{+0.16}_{-0.15}$ &$2.75^{+0.20}_{-0.18}$ &$9.12^{+0.43}_{-0.41}$  &$12.10^{+0.20}_{-0.21}$  &$6.13^{+0.12}_{-0.13}$ \\

FeK$\alpha$ emission line ($\times 10^{-13}$)&$1.02^{+0.80}_{-0.44}$ &$0.79^{+0.21}_{-0.27}$ &$1.15^{+0.26}_{-0.24}$ &$1.00^{+0.32}_{-0.28}$ &$0.71^{+0.46}_{-0.39}$ &$2.24^{+1.07}_{-1.09}$  &$2.78^{+1.04}_{-0.98}$  &$1.61^{+0.47}_{-0.41}$ \\
Neutral reflection$^2$	($\times 10^{-13}$)  &$--$                   & $--$                  &$<4.47$                &$<8.32$                &$3.31^{+3.30}_{-3.30}$ &$<1.23$                 &$<0.04$               &$<0.03$ \\\hline
UV monochromatic$^3$	                     &                       &                       &                       &                       &                       &                        &                      & \\\hline
UVW2 ($\times 10^{-15}$)			         &$2.80\pm0.10$          &$2.03\pm0.08$          &$--$                   &$--$                   &$1.95\pm0.08$          &$6.51\pm0.14$  &$5.08\pm0.14$                  &$3.19\pm0.11$ \\
UVM2 ($\times 10^{-15}$)			         &$2.60\pm0.08$          &$3.19\pm0.05$          &$--$                   &$--$                   &$1.68\pm0.09$          &$5.73\pm0.20$  &$4.29\pm0.18$                  &$--$ \\
UVW1 ($\times 10^{-15}$)			         &$3.50\pm0.05$          &$3.19\pm0.04$          &$--$                   &$--$                   &$2.17\pm0.06$          &$6.07\pm0.13$  &$4.83\pm0.16$                  &$3.17\pm0.13$ \\
U ($\times 10^{-15}$)	       		         &$2.67\pm0.03$          &$3.13\pm0.02$          &$--$                   &$--$                   &$2.39\pm0.06$          &$4.88\pm0.12$  &$3.67\pm0.13$                  &$2.81\pm0.10$ \\\hline

$\rm F_{2\kev}$ ($\times 10^{-12}$)          &$1.12^{+0.05}_{-0.05}$ &$1.51^{+0.04}_{-0.03}$ &$1.78^{+0.04}_{-0.04}$ &$1.59^{+0.04}_{-0.04}$ &$0.63^{+0.03}_{-0.04}$ &$1.91^{+0.09}_{-0.09}$  &$2.49^{+0.09}_{-0.08}$ &$1.29^{+0.04}_{-0.03}$\\
$\alpha_{\rm OX}$                            &$1.228$               &$1.163$               &$--$                   &$--$                   &$1.244$               &$1.231$                &$1.148$              &$1.188$ \\
$\rm \log \rm L_{2-10\kev}$                  &$42.82^{+0.05}_{-0.02}$&$42.96^{+0.02}_{-0.02}$&$43.06^{+0.01}_{-0.01}$&$43.02^{+0.01}_{-0.01}$&$42.64^{+0.03}_{-0.03}$&$43.15^{+0.01}_{-0.01}$ &$43.27^{+0.04}_{-0.03}$ &$42.98^{+0.01}_{-0.01}$ \\
$\lbol$ ($0.001-100\kev$)                    &$43.89^{+0.02}_{-0.02}$&$43.99^{+0.01}_{-0.01}$&$--$                   &$--$                   &$43.65^{+0.02}_{-0.02}$&$44.05^{+0.01}_{-0.01}$ &$44.04^{+0.01}_{-0.01}$ &$43.82^{+0.01}_{-0.01}$ \\
$\lambdaedd$                                 &$0.0130$               &$0.0163$               &$--$                   &$--$                   &$0.0075$               &$0.0187$                &$0.0183$              &$0.0110$ \\\hline

\end{tabular}

$^1$ Unabsorbed power-law flux estimated in the energy range $2-10\kev$.\\
$^2$ The reflected emission due to Compton down scattering of the hard X-ray photons by a neutral medium, as estimated using the model {\it pexrav}.\\
We did not quote this for obs1 and obs2 which are \xmm{} observations and do not cover the above $10\kev$ range.\\ 
$^3$ The UV monochromatic fluxes are measured from \xmm{} OM and \swift{} UVOT instrument. See Table \ref{Table:pheno} for the model fit. The fluxes are in the units of $\funit$\AA{}$^{-1}$.
\end{table*}


\subsection{The Iron K line and the Compton hump}\label{subsec:reflection}

We detected the presence of a weak Fe K$_{\alpha}$ emission line in the source spectra for all observations with a flux of $1.0^{+0.8}_{-0.4}\times 10^{-13} \funit$ for obs1, that remain consistent within 3$\sigma$ uncertainties for all the observations (See Table~\ref{Table:flux}). The iron line is narrow in nature ($\sigma <0.1\kev$) for obs1, obs2, obs3 and obs4. We found an upper limit on the Fe line width of $\sigma < 0.45\kev$ and $<0.44\kev$ for obs5 and obs6 respectively. For obs7 and obs8, due to longer exposure of \nustar{}, we were able to constrain the Fe line width and found $\sigma =0.24^{+0.33}_{-0.16}\kev$ and $0.19^{+0.14}_{-0.10}\kev$ for obs7 and obs8 respectively. In all observations, the addition of {\it pexrav} to the phenomenological set of models did not improve the fit statistic, and we only got an upper limit on the reflection fraction. This result shows no Compton hump present above $10\kev$ in the source spectra. 


\subsection{The power-law, soft excess and UV correlation analysis }

We found the power-law photon index $\Gamma$ for obs1 and obs2 to be slightly steeper than the rest of the observations. Although, the best-fit value of $\Gamma$ remain within errors for the {\it optxagnf} model, it showed a significant variation when we use the model {\it relxill} (e.g., $\Gamma= 1.88^{+0.02}_{-0.08}$ in 2002 and $1.58^{+0.02}_{-0.03}$ in 2021). These results are in consistent with previous studies \citep{2018MNRAS.480.1522L,2020MNRAS.495.3373E}. We have calculated the $2-10\kev$ unabsorbed power-law flux and quoted these values in Table~\ref{Table:flux}. We also estimated the UV monochromatic flux  at $2500\AA$ using the UVW1 band from the optical monitor (OM) and UVOT filter of \xmm{} and \swift{} respectively (See Fig~\ref{fig:variation_flux}). We corrected the source count rates for the Galactic extinction using the CCM extinction law \citep{1989ApJ...345..245C} with a color excess of $\rm E(B-V) = 0.0134$ and the ratio of total to selective extinction of $\rm R_{V}=A_{V}/E(B-V)=3.1$, where, $\rm A_{V}$ is the extinction in the V band. From Table~\ref{Table:flux}, it is evident that both $2-10\kev$ power-law flux and UV monochromatic flux varied significantly between observations. The power-law flux rises from $4.2^{+0.5}_{-0.2}\times 10^{-12}\funit$ in 2002 to $12.1^{+0.2}_{-0.2}\times 10^{-12}\funit$ in 2020 and then declines to $6.1^{+0.1}_{-0.1}\times 10^{-12}\funit$ in 2021. The UV monochromatic flux (UVW1) also rises from $3.50\pm0.05\times 10^{-15}\funit$\AA{}$^{-1}$ in 2002 to $6.07\pm 0.13\times 10^{-15}\funit$\AA{}$^{-1}$ in 2018 and then declines to $3.17\pm0.13\times 10^{-15}\funit$\AA{}$^{-1}$ in 2021. Fig.~\ref{fig:variation_flux} shows the soft excess flux, the power-law flux, the UV monochromatic flux and the Iron line emission flux variation for the last two decades. We notice that the power-law flux and the UV flux follow same temporal pattern. However, the soft excess flux does not follow the trend and shows an unique spectral or flux evolution. 

\subsection{The evolution of the SED }

\begin{figure*}

\centerline{\includegraphics[width=15cm]{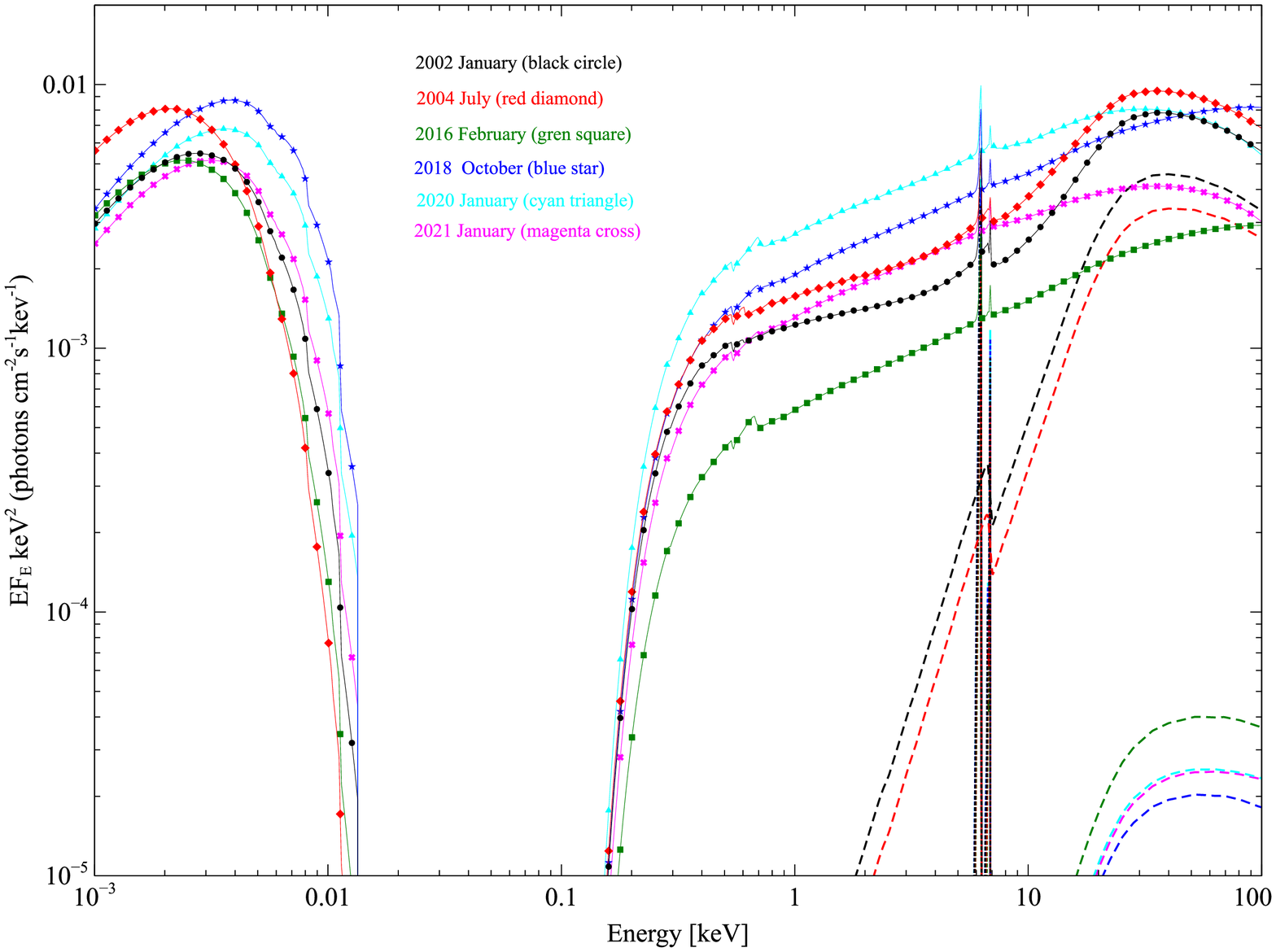}}
\caption{ The best-fit UV/X-ray models for 2002 January (black circle), 2004 July (red diamond), 2016 February (green square), 2018 October (blue star), 2020 January (cyan triangle) and 2021 (magenta cross) derived using 2 \xmm{} and 4 simultaneous \swift{} plus \nustar{} observations. The ionized reflection model describing the soft excess and power-law emission and the optical-UV emission described by the {\it diskbb} are shown for each epoch with different marker and color. The {\it MyTorusL} (dotted lines) and {\it MyTorusS} (dashed lines) components for each epoch are shown with the same colors. 2016 (green square) and 2020 (cyan triangle) observations represent the lowest and highest flux state of the source. The two \suzaku{} observations are not considered due to absence of simultaneous optical-UV data.} \label{fig:sed_variation}

\end{figure*}

We summarize the spectral evolution of Mrk~590 in Fig.~\ref{fig:sed_variation} using the {\it relxill} model as described in Table~\ref{Table:relxill}. The figure shows the best-fitting UV to X-ray broadband models derived from all the observations used in our work, obs1 in black, obs2 in red, obs5 in green, obs6 in blue, obs7 in cyan and obs8 in magenta. The ionized reflection model describing the soft excess and power-law emission and the optical-UV emission described by the {\it diskbb} are shown for each epoch with different marker and color. The {\it MyTorus} model components - the {\it MyTorusL} and {\it MyTorusS} are shown in dotted and dashed lines respectively. Clearly Mrk~590 has shown some unique disk and corona properties over the past few decades. 


\subsection{Estimating the $\lambdaedd$ at different epochs}

We have calculated the bolometric luminosities ($\rm L_{bol}$) of the source at each epoch from our physical broadband spectral modelling in the energy range $0.001-100\kev$. We preferred the ionized reflection model for this purpose which provided a relatively better description of the source spectra at all epochs. Next, we estimated the $\lambdaedd= \rm L_{bol} /L_{Edd}$ during each epoch assuming a black hole mass of $4.5\times 10^{7}\mbh$. We find a sub-Eddington accretion rate for all the observations and are listed in Table~\ref{Table:flux}. However, the values are not consistent between observations and we plotted them ($\lambdaedd$) in Fig.~\ref{fig:variation_flux}. As expected, the value of accretion rate follow a similar trend of variation of that of power-law and UV monochromatic flux. We also plotted the logarithm of the Eddington ratio vs the power-law slope $\Gamma$ at each epoch (Fig.~\ref{fig:correlation}). 


\subsection{$\alpha_{\rm OX}$ vs. $\rm L_{2500\AA}$ correlation}

\begin{figure*}
\centering 
  
\includegraphics[width=8.7cm]{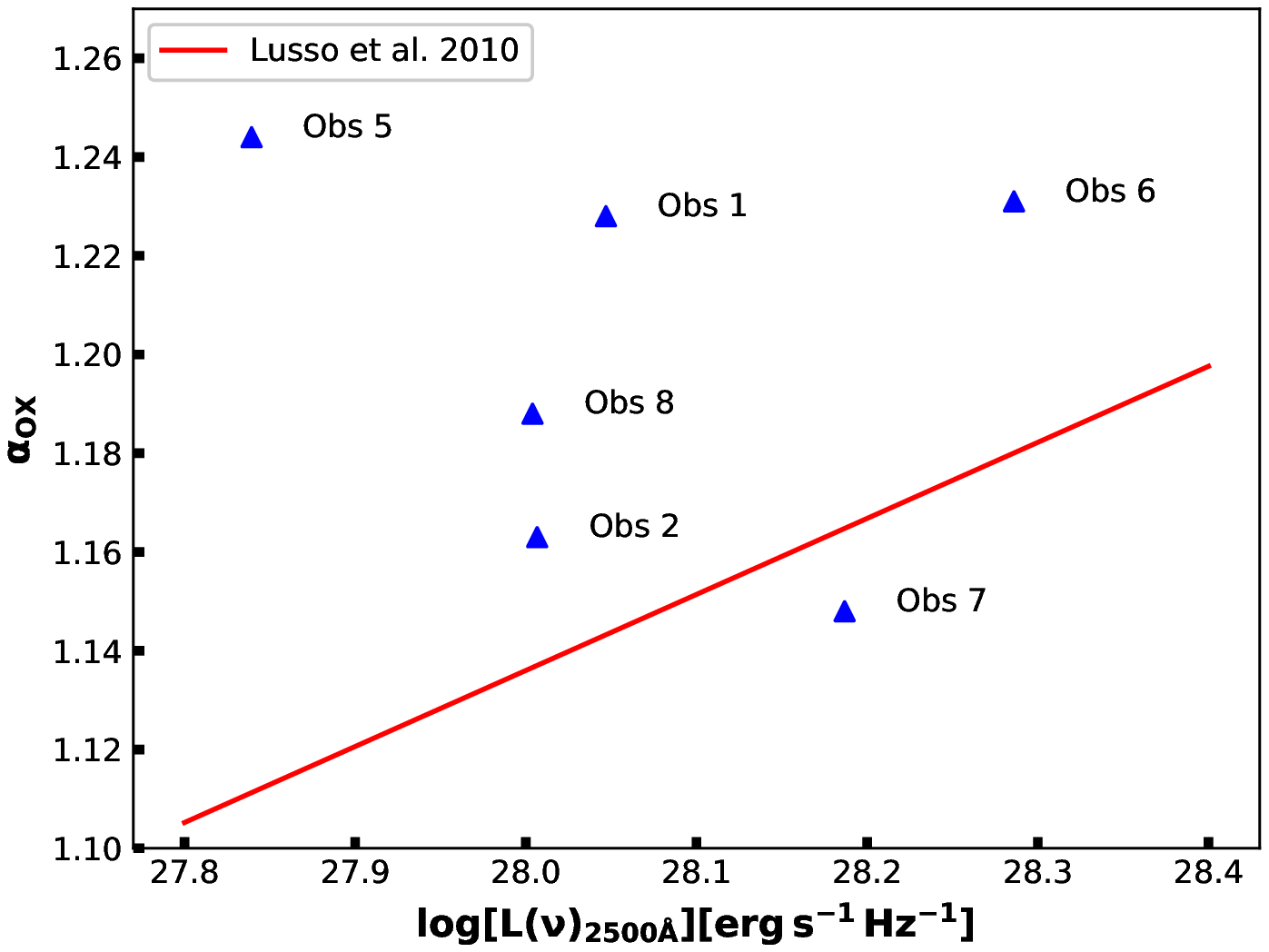}
\includegraphics[width=8.7cm]{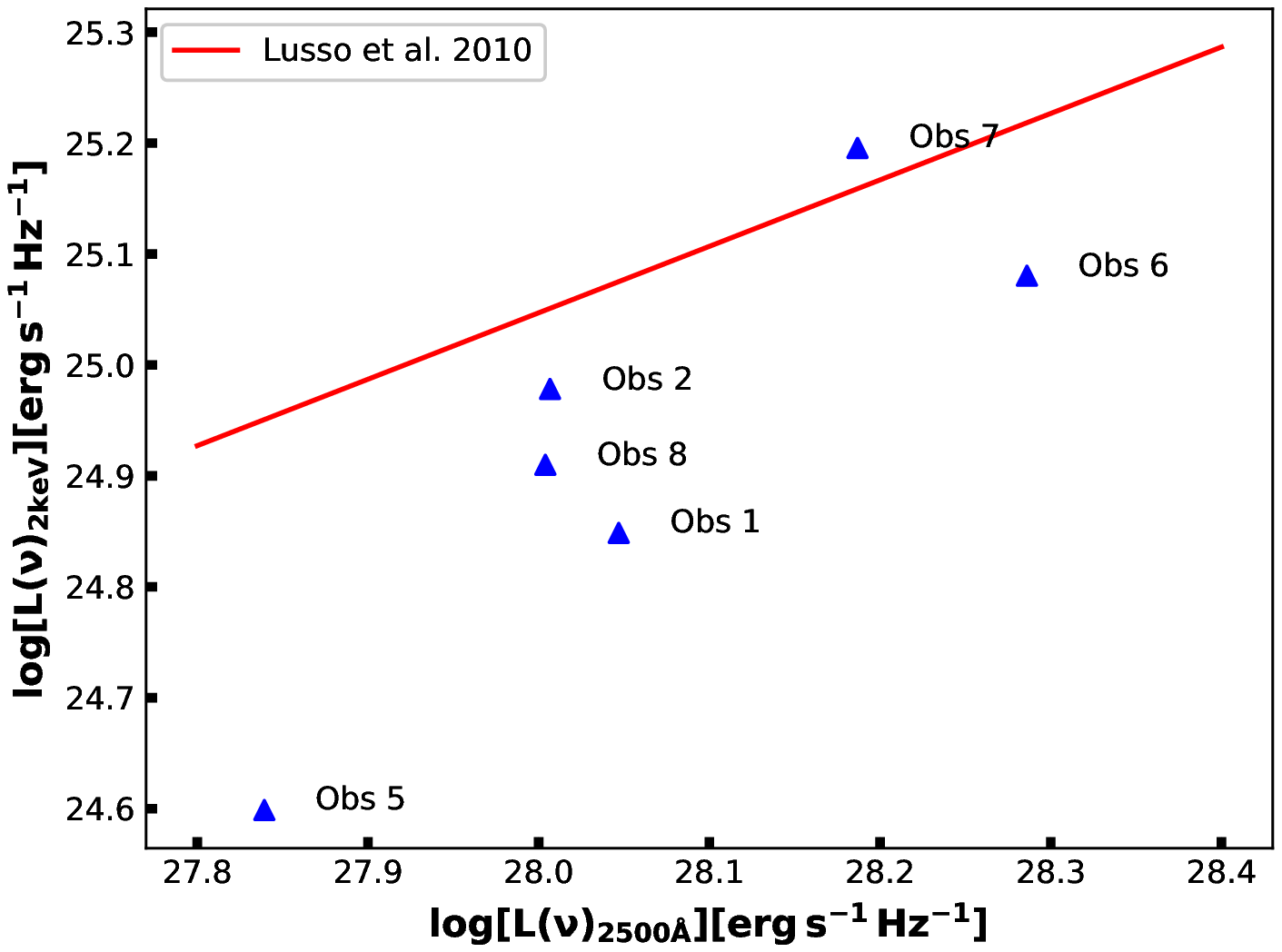}
\includegraphics[width=8.7cm]{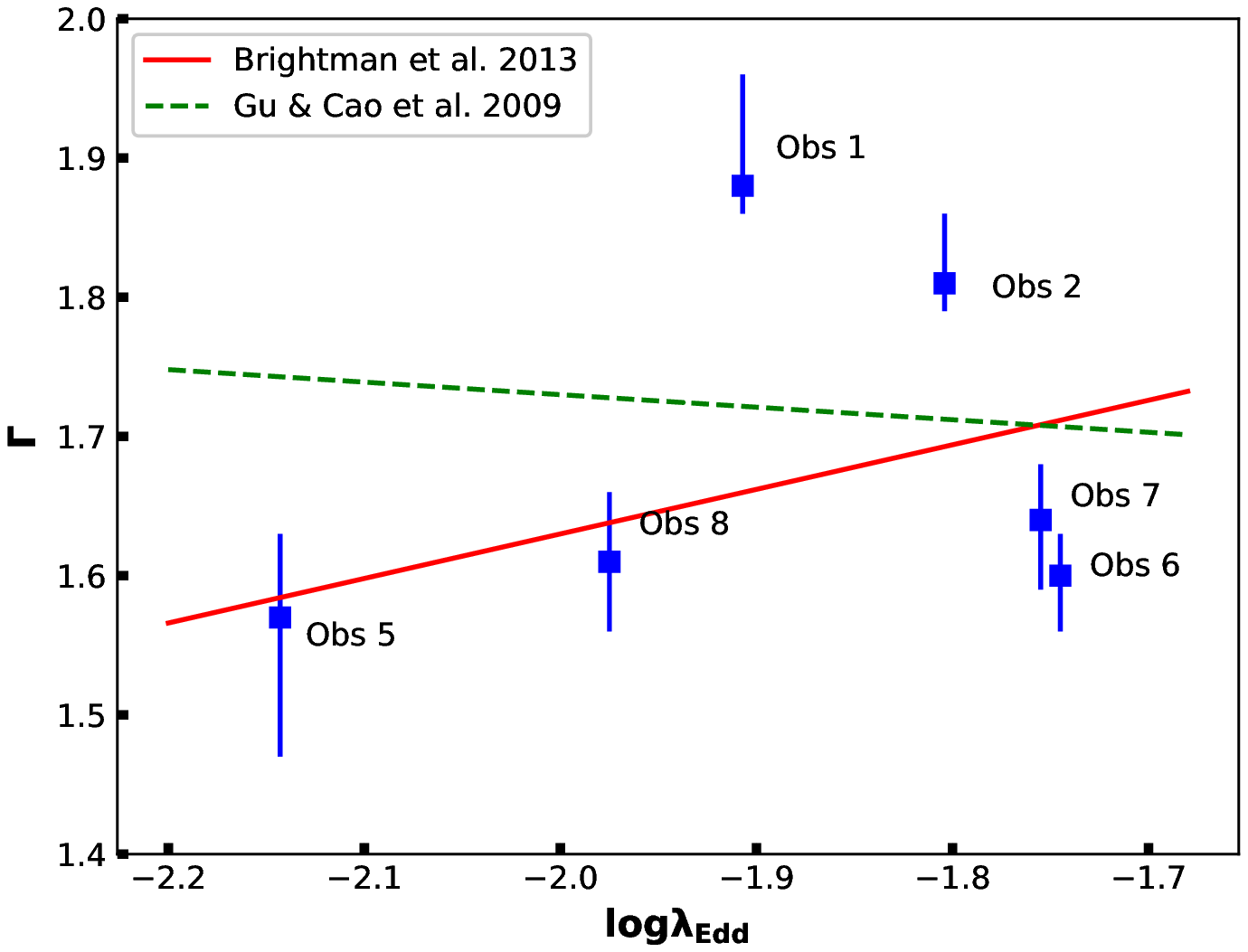}
\includegraphics[width=8.7cm]{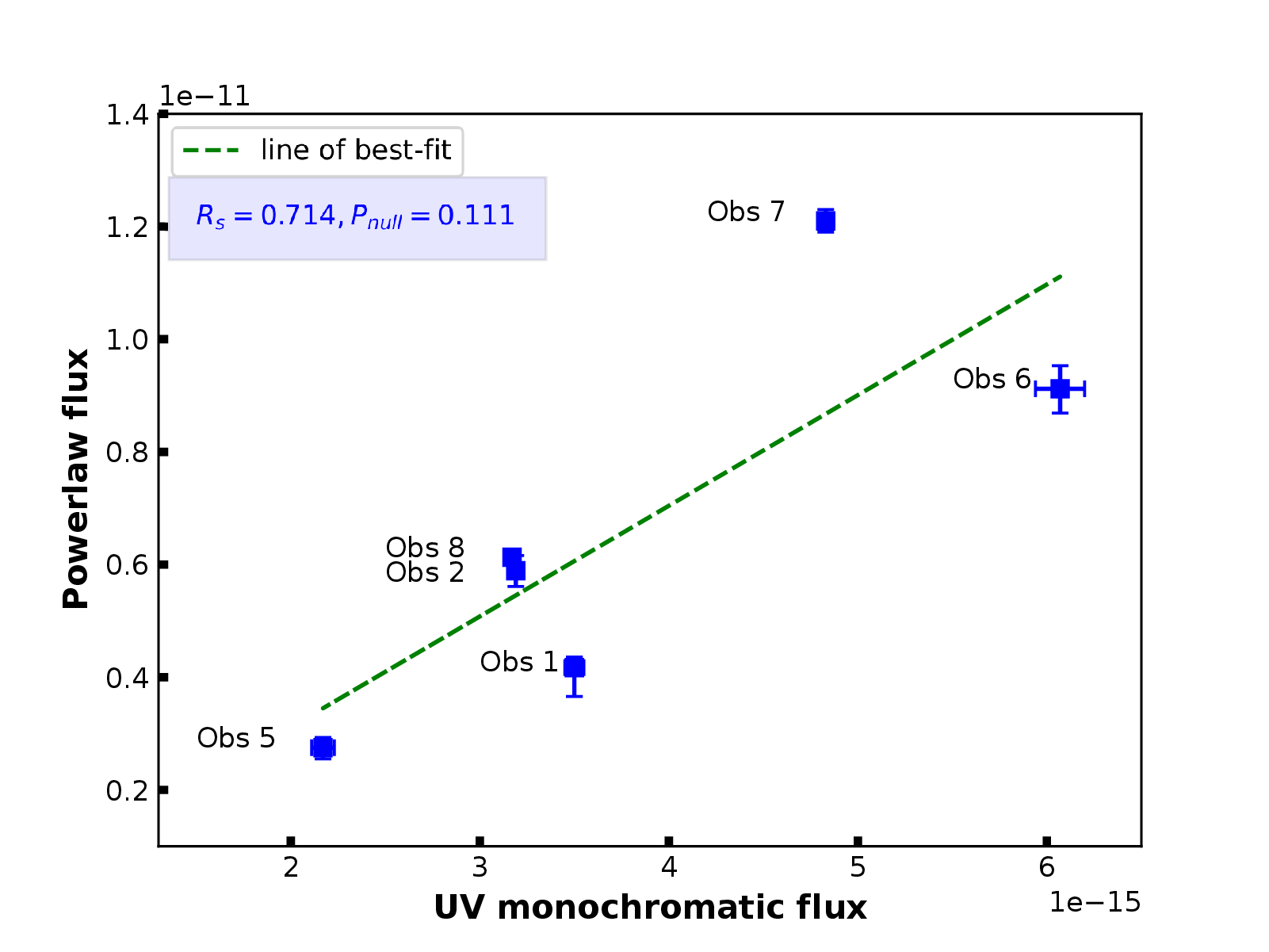}

\caption{{\it Top Left:}The relationship between the 2500\AA{} luminosity and $\alpha_{OX}$ of the changing-look AGN Mrk~590 showing lack of anti correlation. {\it Top Right:} The relationship between the logarithm of the luminosity at $2\kev$ and the UV monochromatic luminosity at 2500\AA{} of the changing-look AGN Mrk~590. {\it Bottom Left:} The relationship between the logarithm of the Eddington ratio vs the power-law slope of the changing-look AGN Mrk~590. {\it Bottom Right:} The correlation plot between the UV monochromatic flux at 2500\AA{} and the $2-10\kev$ power-law flux. We calculated the Spearmans correlation coefficient ($R_{s}=0.714$) and did not find any significant correlation between the two fluxes. The two \suzaku{} observations are not considered as simultaneous optical-UV data is not available.} \label{fig:correlation}

\end{figure*}

We used absorption-corrected UV monochromatic and $2\kev$ fluxes to calculate the $\alpha_{\rm OX} = -0.384 \log[\rm L_{2500 \rm \AA}/L_{2\kev}]$~\citep{1979ApJ...234L...9T}, defined as the power-law slope joining the $2\kev$ and the $2500$\AA{} flux for a given source. The $\alpha_{\rm OX}$ values show variation between observations and quoted in \ref{Table:flux}.  Fig.~\ref{fig:correlation} shows the correlation between the $\alpha_{\rm OX}$ vs $L_{2500 \rm \AA}$ and the $\rm L_{2\kev}$ vs $L_{2500 \rm \AA}$. To compare our results we over plot the best-fit $\alpha_{\rm OX}$ vs $L_{2500 \rm \AA}$ relation found by \citep{2016ApJ...819..154L} and we see that they do not follow this relation.  



\section{Discussion}\label{sec:discussion}

Mrk~590 is well studied both as an individual and part of sample studies in the past~\citep{1993ApJ...414..552O, 2012ApJ...759...63R,2014MNRAS.441.2613L, 2014ApJ...796..134D,2016MNRAS.457.3896L,2018ApJ...866..123M, 2019MNRAS.486..123R,2021MNRAS.502L..61Y}. This source has displayed dramatic changes in amplitude of broad Balmer emission lines between 2006 and 2017. Both \citet{2012ApJ...759...63R} and \citet{2016MNRAS.458.4198M} studied obs3 and obs4 and found that the soft excess vanished in 2012 within seven years, and no relativistic FeK$_{\alpha}$ line was present in the source spectra. In 2015, first time for a CLAGN, very long baseline interferometry (VLBI) observations at 1.6 GHz revealed the presence of a faint ($\sim 1.7 \rm mJy$) parsec scale ($\sim 1.4$ pc) radio jet. Both the changing-look nature of the source and the parsec scale jet has been accredited to the source's variable accretion rate or episodic accretion events. Our work confirms that the soft excess emission has vanished within seven years between 2004 and 2011 and never reappeared. The source spectra showed flux variability in optical-UV, soft and hard X-ray bands. We found a neutral, relatively narrow Fe emission line present in all data sets but no Compton hump above $10\kev$. No relativistic Fe emission line is detected in any of the observations. In the light of these results, we answer the following scientific questions.

\subsection{Origin and nature of the vanishing soft excess}

The soft X-ray excess emission below $2\kev$ is very common in type 1 AGNs, and the origin is still in debate. Our phenomenological set of models revealed the presence of soft excess emission only in obs1 and obs2. Our result is consistent with \citet{2012ApJ...759...63R}. 


We first investigate whether variable obscuration is responsible for this flux variation. An obscuring cloud crossing the line of sight and causing changes in observed light curves \citep{1989ApJ...340..190G,2016ApJ...826..186G} may cause the soft excess to vanish. However, we did not find the presence of intrinsic absorption in any of the source spectra. This clearly rules out variable obscuration to be the reason behind this vanishing soft excess. Next, we discuss the results of two physical models used in our work to find the origin of the soft excess emission.

In the thermal Comptonization model {\it optxagnf}, soft excess emission is produced due to thermal Comptonization of disk optical-UV photons by a warm ($kT \sim 0.1-0.2\kev$) optically thick ($\tau \sim 10-20$) corona surrounding the inner regions of the disk. Hence the vanishing soft excess requires a vanishing warm corona or simply a significant change in the size of the warm Comptonizing medium ($\rm r_{cor}$). Here, the disk makes transitions between a cold$+$warm disk and a cold disk. This excess energy generation in the innermost disk must be provided by the increase in accretion rate, which could not all be released in the form of radiation. This excess energy raises the temperature and pressure and transforms the innermost accretion disk to the warm Comptonization medium. So, the absence of soft excess should follow with a decrease in the accretion rate \citep{2012mnras.420.1848d,2022ApJ...925..101T}. If this scenario is true, there must be a correlation between the soft excess flux, optical-UV flux, and mass accretion rate. In addition, the spectral state transition due to disk evaporation and change in the accretion rate, leads to spectral hardening of the photons arising from the hot corona.

\begin{figure}
  \centering 
  
\includegraphics[width=8.5cm]{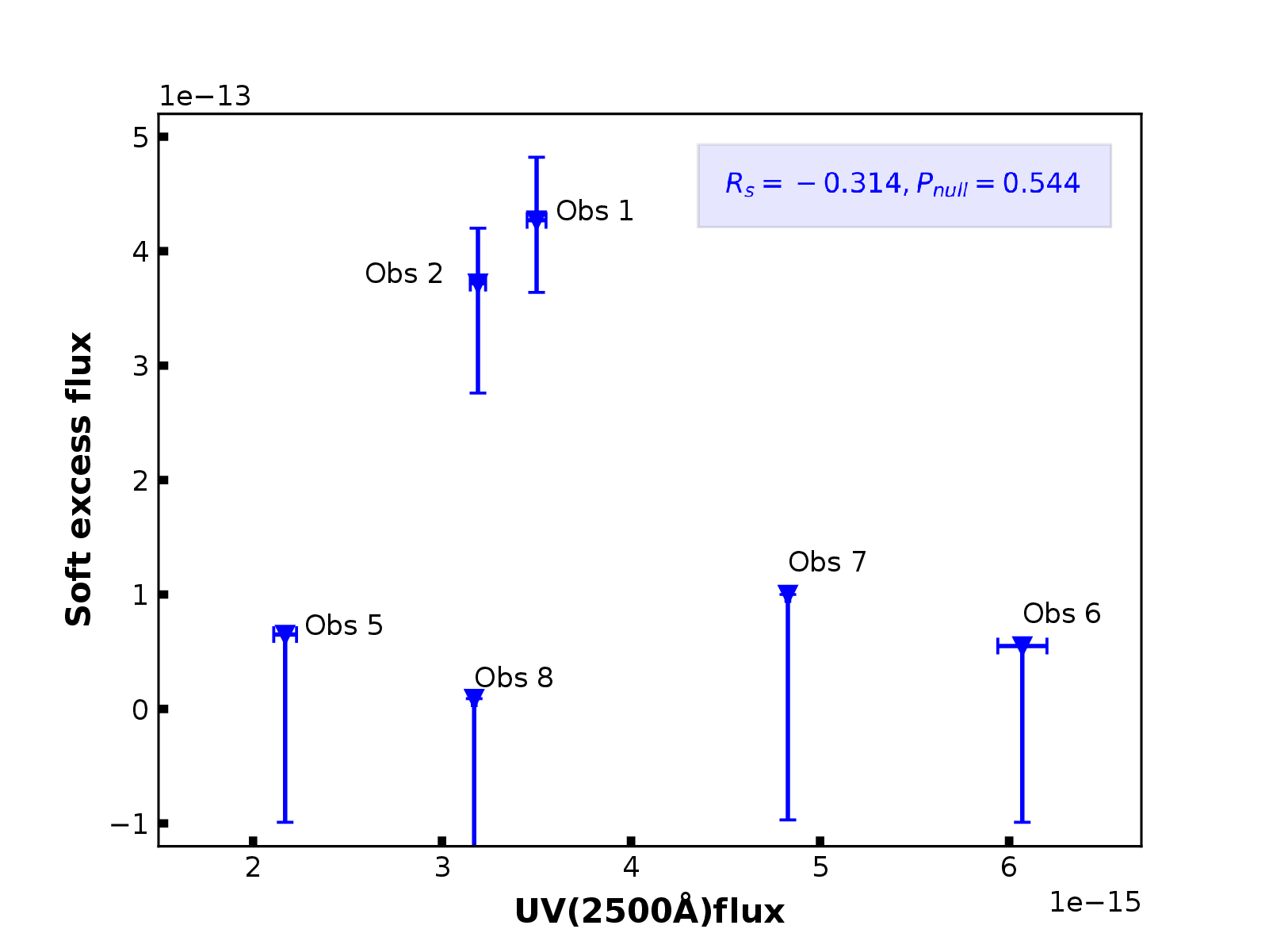}

\caption{The correlation plot between the $0.3-2.0\kev$ soft excess emission flux and the UV monochromatic (2500\AA) flux. We do not see any significant correlation ($R_{s}=-0.314$). The two \suzaku{} observations are not considered as simultaneous optical-UV data is not available.} \label{fig:correlation_soft_uv}

\end{figure}
\begin{figure}
  \centering 
  
\includegraphics[width=8.5cm]{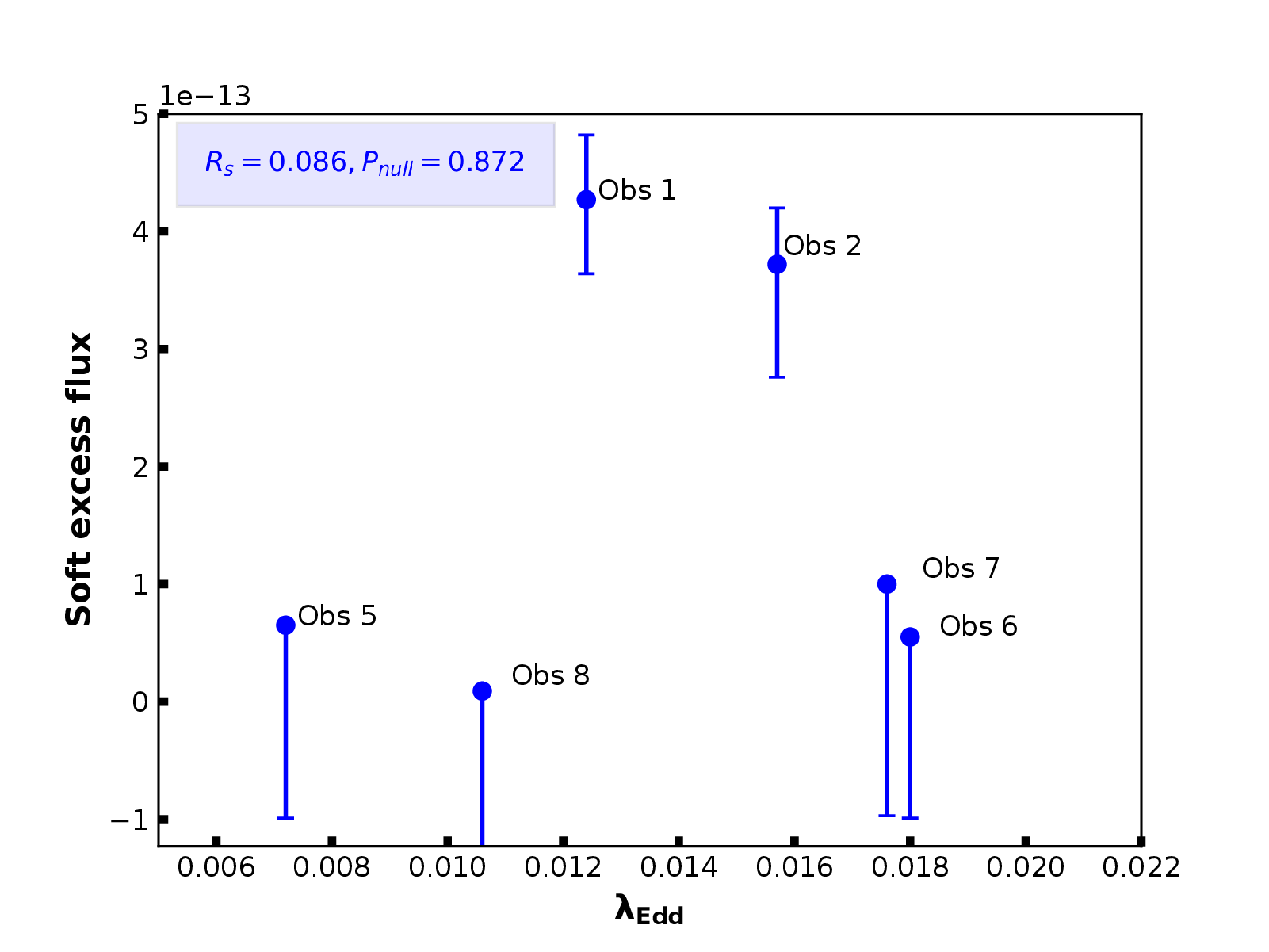}

\caption{The correlation plot between the $0.3-2.0\kev$ soft excess emission flux and the accretion rate, $\lambda_{\rm Edd}$ for all six observations where simultaneous optical-UV data is available. Hence, the two \suzaku{} observations are not considered. The figure shows no significant correlation between the two fluxes ($R_{s}=0.086$).} \label{fig:correlation_soft_edd}

\end{figure}

However, in Mrk~590, {\it optxagnf} needed an additional {\it diskbb} component to model the UV bump in obs1, obs2, and obs5 as the model overestimated the optical-UV flux when extrapolated from the spectral fitting of the X-ray energy band. Now, if the soft excess is very weak or absent, the warm corona responsible for the soft excess may instead contribute to the optical-UV band. But, we found that we did not require this {\it diskbb} component for obs6, obs7, and obs8, where soft excess is absent. The improvement in fit-statistics after the addition of the {\it diskbb} component was significant. Table~\ref{Table:flux} shows that UV monochromatic flux measured for obs1, obs2, and obs5 is significantly lower than UV flux measured during obs6, obs7, and obs8. In Mrk~590, we see the soft excess flux drop four times within seven years. We expect an increase in the value of $\rm r_{cor}$ and a decrease in the accretion rate as found in other CLAGNs, e.g., Mrk~1018~\citep{2018MNRAS.480.3898N} and NGC~1566~\citep{2022ApJ...925..101T}. In NGC~1566, the soft excess flux component decreases by a factor $>45$, and a significant change in the size of the warm Comptonizing medium ($\rm r_{cor}$) is found where $\rm r_{cor}$ increased from $\sim 26\rm r_{g}$ during high flux state in 2015 to $50\rm r_{g}$ during low flux state in 2018. However, for Mrk~590, we could not determine the exact size ($\rm r_{corona}$), electron temperature, or optical depth ($kT_{e}$ and $\tau$) of the Comptonizing corona even when the soft excess is present. We did not find any significant decrease in luminosity and accretion rate. In addition, the {\it optxagnf} best-fit (Table~\ref{Table:optxagnf}) shows no significant change in power-law $\Gamma$ or $\rm f_{pl}$, indicating no spectral hardening between observations. The results indicate no correlation between the optical-UV and soft excess flux. Fig.~\ref{fig:correlation_soft_uv} and Fig.~\ref{fig:correlation_soft_edd} show this lack of correlation, where we plot the soft excess versus the UV flux and the accretion rate, respectively. We also found that {\it optxagnf} could not model the observed exponential cut-off in power-law continuum in obs7 and obs8. The source spectra did not require black hole spin to model the soft excess emission. This result is consistent with Mrk~1018 and NGC~1566 but contradicts the recent findings in other type 1 AGNs where soft excess is present~\citep{2019ApJ...871...88G,2020MNRAS.497.4213G, 2021ApJ...908..198G, 2021ApJ...913...13X}. These results show that the thermal Comptonization of disk photons, that successfully explained the soft excess flux variation in other CL-AGN such as NGC 1566 and Mrk 1018, is unable to explain the vanishing soft excess and high energy cut-off observed in Mrk~590.


The ionized reflection model {\it relxill} provided a good fit for all the observations. In this model, the untruncated accretion disk approaches the innermost stable circular orbit due to high black hole spin. The hard X-ray photons from the corona illuminate the disk, ionize it, and emits fluorescent emission lines. These lines are blurred due to extreme gravity near SMBH. In this scenario, the power-law flux and the soft excess flux should have a strong correlation between them. So, the decrease in soft excess flux may occur due to changes in disk and corona properties. Some possibilities are (a) the disk becoming a truncated one ($r_{in}>50r_{g}$) due to disk evaporation or (b) becoming highly ionized ($\xi\sim10^{4}\xiunit$)\citep{2005MNRAS.358..211R}. The disk may become highly ionized due to spectral hardening as harder illuminating spectra have greater ionizing power. But spectral hardening will also give rise to a strong and broad Fe emission line and a Compton hump. The change in soft excess strength should also affect the reflected flux or the reflection fraction that determines the ratio of intensity emitted towards the disk compared to escaping to infinity.

\begin{figure}
  \centering 
  
\includegraphics[width=8.5cm]{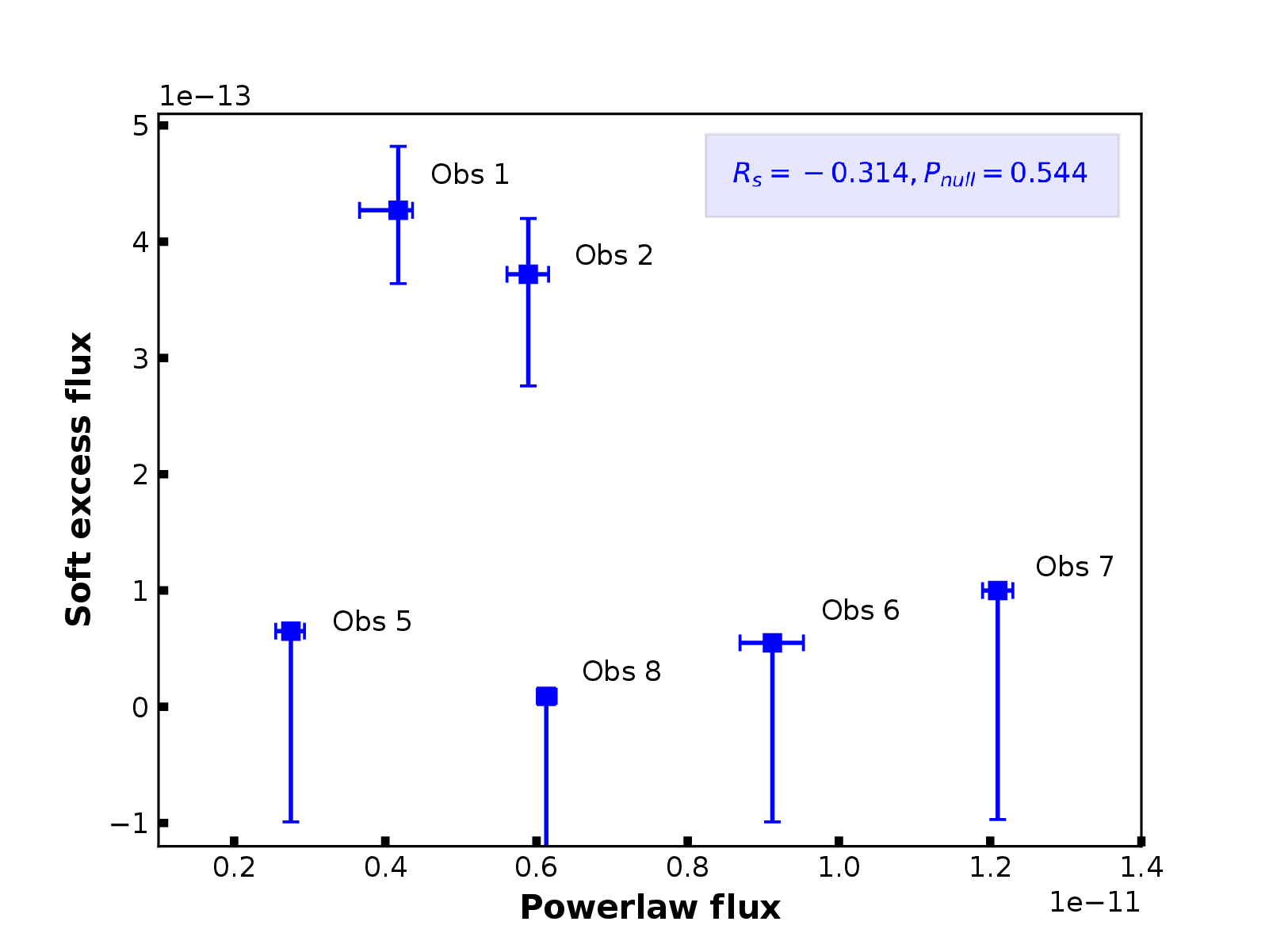}

\caption{ The correlation plot between the $0.3-2.0\kev$ soft excess emission flux and the $2-10\kev$ power-law flux. We did not find any significant correlation between the two fluxes ($R_{s}=-0.314$).} \label{fig:correlation_soft_po}

\end{figure}

We note that Mrk~590 does not fit this description. The soft excess in obs1 and obs2 were described by a non-rotating black hole ($r_{in}$ fixed at $6 r_{g}$) and a slightly ionized accretion disk. This result contradicts recent studies of other type 1 AGNs, where the reflection model favors a rotating black hole and the inner part of the disk approaches the inner-most stable circular orbit ($1.25 r_{g}$) \citep{2019ApJ...871...88G,2020MNRAS.497.4213G, 2021ApJ...908..198G}. We note a significant increase in disk ionization between observations (from $\log\xi=0.52^{+0.77}_{-0.30}$ in obs2 to $\log\xi \sim 3$ in obs3). We also found a significant decrease in $\Gamma$ value from obs1 to obs8, indicating a spectral hardening. For obs7 and obs8, we found a well constrained, high-energy cut-off of $92^{+55}_{-25}\kev$ and $60^{+10}_{-8}\kev$ respectively, better described by the {\it relxill} model. Although these values are relatively lower compared to other Seyfert 1s ($200-300\kev$) \citep{2016MNRAS.456..554G,2017ApJS..233...17R, 2017MNRAS.467.2566F,2021A&A...655A..60A}, similar low values of $\rm E_{cut}$ have been found in recent sample studies of \swift{}/BAT selected AGNs~\citep{2022ApJ...927...42K}. This low energy cut-off may indicate a decrease in the plasma temperature of the corona only if it was higher during obs1 and obs2. But we can not test this scenario due to the non-availability of data beyond $10\kev$. Now hard X-ray photons illuminate the disk. So less energetic photons mean low illumination. However, we note that the disk is still moderately ionized and is capable of \citep{2005MNRAS.358..211R} producing fluorescent emission lines. More importantly, we did not find any broad Fe emission line or Compton hump in any of the source spectra. The reflected flux and reflection fraction does not show statistically significant variation and remain within the $3\sigma$ value. When we plotted the soft excess flux versus the power-law flux ($2-10\kev$), we did not find any significant correlation (See Fig.~\ref{fig:correlation_soft_po}). This result is consistent with \citep{2016A&A...588A..70B}, where the shape of reflection at hard X-rays stays constant when the soft excess varies, showing an absence of a link between reflection and soft excess. In Mrk~590, the power-law, the Fe line emission, and UV monochromatic flux follow the same temporal pattern (Fig.~\ref{fig:variation_flux}). This result suggests that the disk and corona are most likely evolving together. But, the soft excess is not responding to this change in disk-corona properties. Hence, the soft excess emission observed in Mrk~590 is not due to ionized reflection from the disk.

So, we have two possibilities. Either we can not distinguish the differences in the ionized reflection and warm Comptonization models due to low-quality data, or these models cannot describe the vanishing soft excess feature observed in Mrk 590. Next, we discuss in detail the possibility of change in accretion disk profile behind the spectral variability in Mrk~590 as found in other CLAGNs.

\subsection{Changing Look nature due to change in disk profiles?}

Previously, state change due to disk evaporation or condensation associated with a factor $2-4$ decrease or increase in luminosity, significant mass accretion rate change, or the combination of both has been suggested as the reason behind the changing-look nature of Mrk~590 \citep{2018MNRAS.480.3898N,2018ApJ...866..123M,2021MNRAS.502L..61Y}.
The soft excess emission in Mrk~590 vanished within seven years. This timescale puts an upper limit on the position of the reprocessing material within seven light-years or roughly 2 parsecs. 
This distance is significantly large compared to the distance (10-100 light days) of the broad-line region (BLR) from SMBH in a typical AGN but comparable to the distance of torus($\sim$ few parsecs). The BLR region in Mrk~590 also has gone through some dramatic changes as after $\sim 10$ years of absence, the optical broad emission lines of Mrk 590 have reappeared~\citep{2019MNRAS.486..123R}. The absence of soft excess even when the source changed its type suggests a lack of correlation between the two phenomena. For further investigation, we study the disk instability in Mrk~590 that may cause the observed flux variation.\\

In Mrk~590, we see a drop in total accretion luminosity from $9.8\times 10^{43}\lunit$ (in obs2) to $4.4\times 10^{43}\lunit$ (in obs5), which then again rise to $1.2\times 10^{44}\lunit$ (in obs6) over a timescale of $\sim14$ years. The amplitude of bolometric drop requires a change in either mass accretion rate or efficiency (or both) of the accretion flow. But the observed timescale ($\sim14$ years) is too fast for the mass accretion rate to change through a standard disk (viscous timescale) by many orders of magnitude \citep{2020ApJ...898L...1R, 2019ApJ...887...15W, 2018MNRAS.480.3898N}  and poses a problem for any standard disk model. If we compare the three timescales for a standard thin disk, the dynamical timescale is the fastest, then the thermal timescale, and then the viscous. 
To compare the disk variability timescale in Mrk~590, we calculate the accretion disk timescales at the inner radius of the disk obtained from the best-fit broadband spectral model. The dynamical ($t_{dyn}$), thermal ($t_{th}$), and viscous ($t_{vis}$) timescales of the accretion disk are given by, \citep{2006ASPC..360..265C}\\

\begin{equation}
    t_{dyn} = \Big(\frac{R^{3}}{GM_{BH}}\Big)^{1/2}
\end{equation}
\begin{equation}
    t_{th} = \frac{1}{\alpha}t_{dyn}, 
\end{equation}
and
\begin{equation}
    t_{vis} \sim \frac{1}{\alpha}\Big(\frac{R}{H}\Big)^{2} t_{dyn},
\end{equation}

where R is the radial distance in the disk, $\alpha$ is the viscosity parameter, and H is the height of the disk. To estimate these values, we first calculate the accretion disk temperature of $ \sim 1.23 \ev$ for an inner disk radius of $ \sim 100R_{g}$, accretion rate of $\mdot \sim 0.018$, and a black-hole mass of $4.75\times 10^7\msol$. This implies $\rm H/R = c_{s}/V_{\phi} \sim 3.6\times10^{-4}$ (where $c_{s} = \sqrt{kT/m_{p}}$ is the sound speed and $v_{\phi} = \sqrt{GM_{BH}/R}$ is the Keplerian orbital velocity). Assuming $\alpha = 0.1$, we finally estimated the dynamical, thermal, and viscous timescales to be $t_{dyn}=$ 3.7days, $t_{th}=$ 37days, and $t_{vis}=$ $7\times 10^5$ year, respectively. So the timescale (7 year) of flux variability in Mrk~590 is much smaller compared to the viscous timescale but longer than the dynamical and thermal timescales at an inner radius of $100R_{g}$. If we consider the sound crossing time of $t_{s} \sim 100R_{g}/c_{s} \sim 20$ years is still an order of magnitude higher than the changing-look time of Mrk~590. So the flux variability in Mrk~590 is not likely due to pressure instabilities in the disk.  However, only if we consider an untruncated thin accretion disk up to $10r_{g}$ then, the variability timescales become much shorter and comparable to the observed timescale in Mrk~590. Similar procedures mentioned above estimate the timescales to be $t_{dyn}=$ 2.84hrs, $t_{th}=$ 28hrs, and $t_{vis}=$ $4\times 10^3$ years, respectively. The sound crossing time becomes $t_{s} \sim 10R_{g}/c_{s} \sim 1$ year. Following this assumption we have an inner disk temperature of $\sim 6\ev$, which is inconsistent with the disk temperature we got from our spectral best-fit using ionized reflection model (See table~\ref{Table:relxill}). Also, the disk this close to SMBH will affect the Keplerian frequency \citep{2001PASJ...53....1K} and should give rise to stronger reflection features in the source spectra, common in other Type 1 AGNs but absent in our spectral analysis. This indicates towards a possible disk truncation above $\rm 10r_{g}$. In addition, the change in accretion profile should affect the accretion rate in Mrk~590. \cite{2019MNRAS.486..123R} found that the broad Balmer broad emission lines in Mrk~590 have reappeared in October 2017, within a time-scale of decades. A similar behaviour has also been observed in Mrk 1018 \citep{2016A&A...593L...8M,2016A&A...593L...9H}. We note that this reappearance of Balmer lines coincide with the increase in the mass accretion rate (See Table~\ref{Table:flux} and Fig.~\ref{fig:variation_flux}). But, when we plot the soft excess flux and $\lambdaedd$, (Fig~\ref{fig:correlation_soft_edd}), we did not find any correlation between these two parameters as previously suggested by \citet{2018MNRAS.480.3898N,2018ApJ...866..123M,2021MNRAS.502L..61Y}. Instead, we found a relatively higher Eddington ratio even when the soft excess is not present. We also note that our best-fit accretion rate ($\sim 0.02\pm0.01$) is consistent with previous studies ($\sim 0.03\pm0.01$) \citep{2018MNRAS.480.1522L}. So the origin soft excess emission in Mrk~590 is likely to be not related to the change in the accretion rate.

In Mrk~590, the observed UV and power-law flux variability follow the same temporal pattern. The corona cools down, and the disk becomes more ionized between observations. Hence a change in the nature of the accretion disk and corona is evident, but not the fundamental process through which the disk-corona evolves. To investigate further, we studied the relation between the $2-10\kev$ power-law slope, $\Gamma$, and the Eddington ratio ($\lambdaedd$) at each epoch. This exercise helps us check how efficiently the disk photons are coupled with the hot corona and how efficient the central engines are. A strong coupling between $\Gamma$ and $\lambdaedd$ implies that a higher accretion rate cools off the corona faster, leading to steeper power-law slopes \citep{1995MNRAS.277L...5P}. Previously \citet{2013ApJS..207...19B} and \citet{2017MNRAS.470..800T} studied a sample of radio-quiet AGN and a BAT-selected AGN sample, respectively, and found a strong correlation between the $\Gamma$ and the $\lambdaedd$. \citet{2009MNRAS.399..349G} investigated the relation for a sample of 57 low-luminous AGN (LLAGN) in the local Universe and found that they follow an anti-correlation. This contradiction suggests the possibility of two modes of accretion above/below some critical transition value of $\lambdaedd$. From Fig.~\ref{fig:correlation}, we find that Mrk~590 does not show any such strong correlation or anti-correlation between the spectral slope and the Eddington rate. This is consistent with \citet{2018ApJ...868...10L}, who did not find any such strong correlation in a sample of low-luminous QSOs. These results clearly show that the disk/corona interaction in Mrk~590 does not follow the typical disk-corona properties of Seyfert 1 AGNs and has unique characteristics. In this context we note that there has been a recent discovery of a changing look phenomenon in an AGN 1ES~1927+654 \citep{trakhtenbrot2019,ricci2020,Laha1ES} the origin of which is still debated. However, the radio, optical, UV and X-ray observations point towards an increase of accretion probably due to magnetic flux inversion, as the primary cause of this event \citep{scepi2021,Laha1ES}.

\subsection{The complex reflection in Mrk 590}

The relativistic reflection from the ionized accretion disk can not explain the spectral variability in Mrk~590. The soft excess flux variation does not correlate with the power-law continuum flux. The lack of a broad Fe emission line in the spectra indicates that relativistic reflection does not dominate the source spectra. We were unable to constrain the Fe abundance of the disk, which is previously observed in other Seyfert 1 AGNs \citep{2009Natur.459..540F,2012MNRAS.422.1914D, 2018ASPC..515..282G, 2020MNRAS.497.4213G,2021ApJ...915...93L} as well. A narrow Fe emission line in the X-ray spectra suggests a distant neutral reflection of the hard X-ray continuum from the outer part of the disk or torus. In Mrk~590, we found that both Fe line emission and the power-law flux follow the same temporal pattern, supporting the idea that narrow Fe line emission is most likely due to a neutral reflection of hard X-ray photons from the outer part of the disk. However, we do not see any Compton hump which arises due to Compton down-scattering of high energy photons by the cold disk or torus. This result is inconsistent with the typical neutral reflection observed in the X-ray spectra of type 1 AGNs. The reflection fraction value was within $3\sigma$ significance throughout observations, and we did not find any change in the disk properties except for the ionization. These results indicate a complex reflection scenario that does not follow the typical disk-corona interaction in reflection-dominated type 1 AGNs.


\section{Conclusions}\label{sec:conclusion}

\begin{itemize}
    \item The soft X-ray excess emission in Mrk~590 vanished within seven years (from 2004 to 2011) and never reappeared in later observations.
    
    \item The power-law $\Gamma$ showed a spectral hardening ($\Gamma=1.88^{+0.02}_{-0.08}$ and $\Gamma=1.58^{+0.02}_{-0.03}$ in 2002 and 2021 respectively) in 19 years.
    
    \item A high-energy cut-off of the power-law component was found in the latest \nustar{} observations ($92^{+55}_{-25}\kev$ and $60^{+10}_{-08}\kev$ for obs7 and obs8 respectively). 
    
    \item We find that the disk becomes more ionized (from $0.52^{+0.77}_{-0.30}$ in 2004 and to $3.30^{+0.55}_{-0.99}$ in 2021) when the soft excess is absent.
    
     \item A neutral Fe$K_{\alpha}$ line emission is detected in all data sets and the line emission flux is almost consistent ($<3\sigma$) between observations. However, no Compton hump was detected in any of the observations.
    
    \item The soft excess flux variability does not correlate with changes in power-law or UV flux observed during these observations. 
    
    \item Mrk~590 showed a sub-Eddington accretion rate ($\lambdaedd=0.01-0.02$) and the soft excess flux has no correlation with Eddington ratio. The accretion rate and inner disk temperature ($1-2\ev$) indicates a disk truncation above $\rm 10r_{g}$.
    
    \item The ionized disk reflection model provided a relatively better description of the source X-ray spectra where the high energy cut-off are found (obs7 and obs8).
    
    \item The warm Comptonization model needed additional disk component to describe the UV bump when the UV flux was low (obs1, obs2 and obs5) and we were unable to constrain the warm corona properties without applying this additional `diskbb' component.
    
    \item Although we get statistically good fit for both the soft excess models, given the data quality, the ionized disk reflection and warm Comptonization models for certain observations do not conform with typical AGN scenario and are not adequate to describe the soft excess feature observed in Mrk 590.
    
    \item The disk instability timescale ($\sim 20$ years) is unable to explain the observed soft excess variation in Mrk~590, making the fundamental process unclear through which the accretion disk evolves.
    
\end{itemize}


\section{Acknowledgements}
The authors are grateful to the anonymous referee for insightful comments which improved the quality of the paper.
RG acknowledges the financial support from IUCAA. This research has made use of the NuSTAR Data Analysis Software (NuSTARDAS) jointly developed by the ASI Science Data Center (ASDC, Italy) and the California Institute of Technology (USA). The results are based on observations obtained with XMM-Newton, an ESA science mission with instruments and contributions directly funded by ESA Member States and NASA. This research has made use of the XRT Data Analysis Software (XRTDAS) developed under the responsibility of the ASI Science Data Center (ASDC), Italy. This research has made use of data obtained from the Suzaku satellite, a collaborative mission between the space agencies of Japan (JAXA) and the USA (NASA).

\section{Data availability}

This research has made use of archival data of \suzaku{}, \swift{}, \nustar{} and \xmm{} observatories through the High Energy Astrophysics Science Archive Research Center Online Service, provided by the NASA Goddard Space Flight Center.

\appendix

\section{The soft excess variability of Mrk 590 at different epochs}\label{appendix_softexcess}

\clearpage
    \begin{figure}
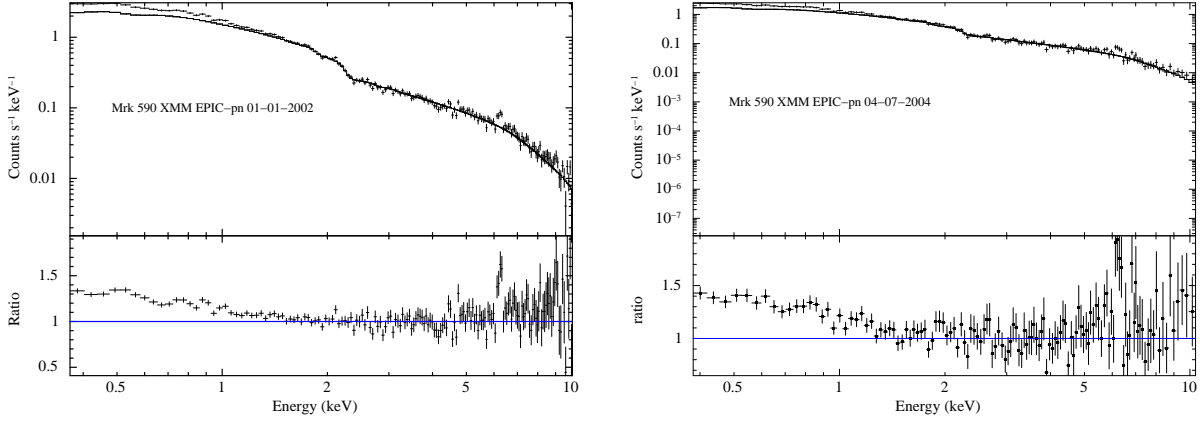

    \centering 

    \hbox{
    \includegraphics[width=5.5cm,angle=-90]{check.excess.mrk590.xmm.4july2004.eps}
    \includegraphics[width=5.5cm,angle=-90]{check.soft.excess.xmm.1stJan.2002.eps}}
    \caption{{\it Left:} The $2.0-5.0\kev$ \xmm{} spectra, obs1 (on left) and obs2 (on right) of Mrk~590 fitted with an absorbed power-law and the rest of the energy band ($0.3-10.0\kev$) extrapolated. The broadband residuals from the fit, showing the presence of soft X-ray excess and a Fe emission line for the two \xmm{} observations. The X-axis represents observed frame energy.} \label{fig:check_excess_obs1_2}

    \end{figure}

    \begin{figure}
    \centering 

    \hbox{
    \includegraphics[width=5.5cm,angle=-90]{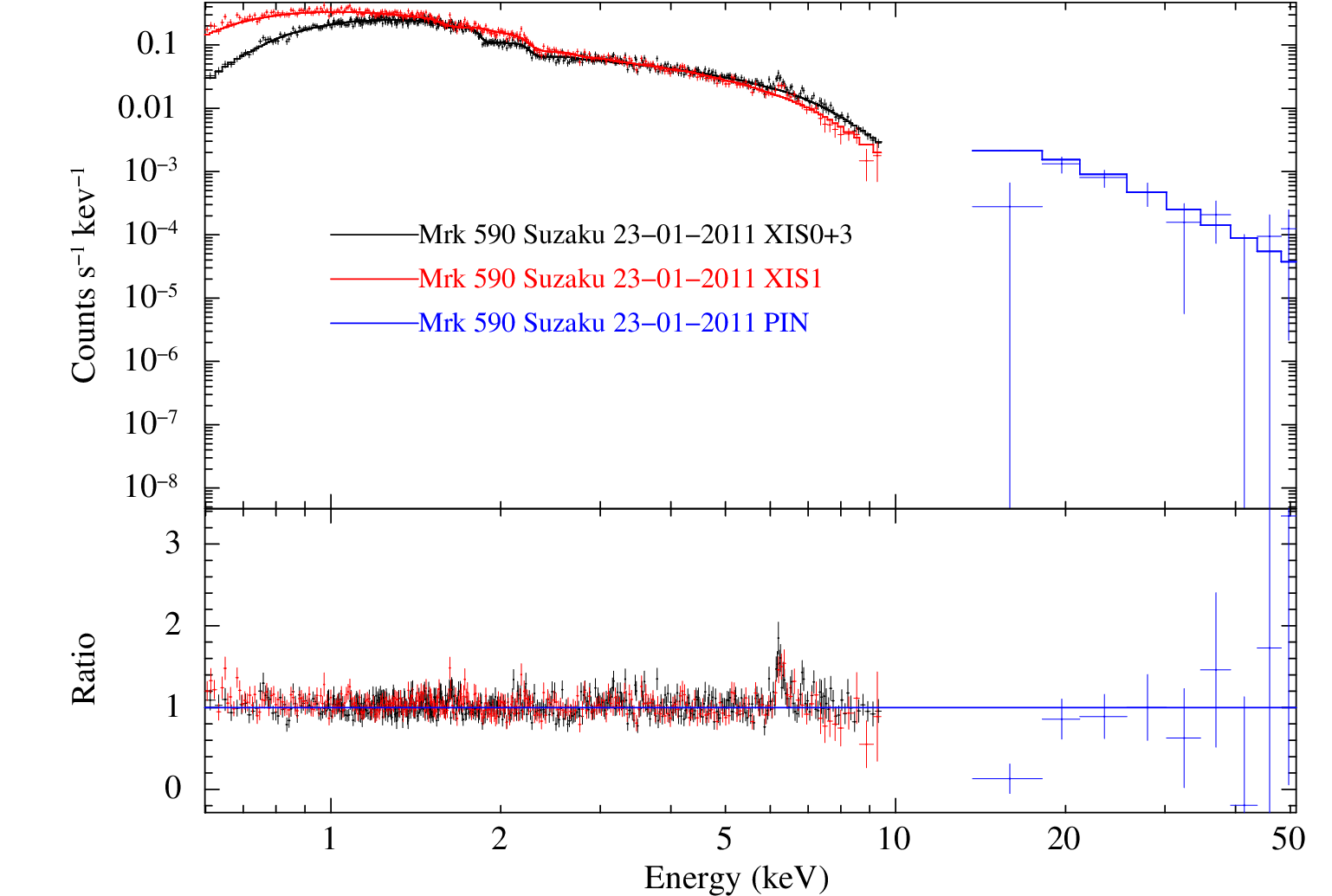}
    \includegraphics[width=5.5cm,angle=-90]{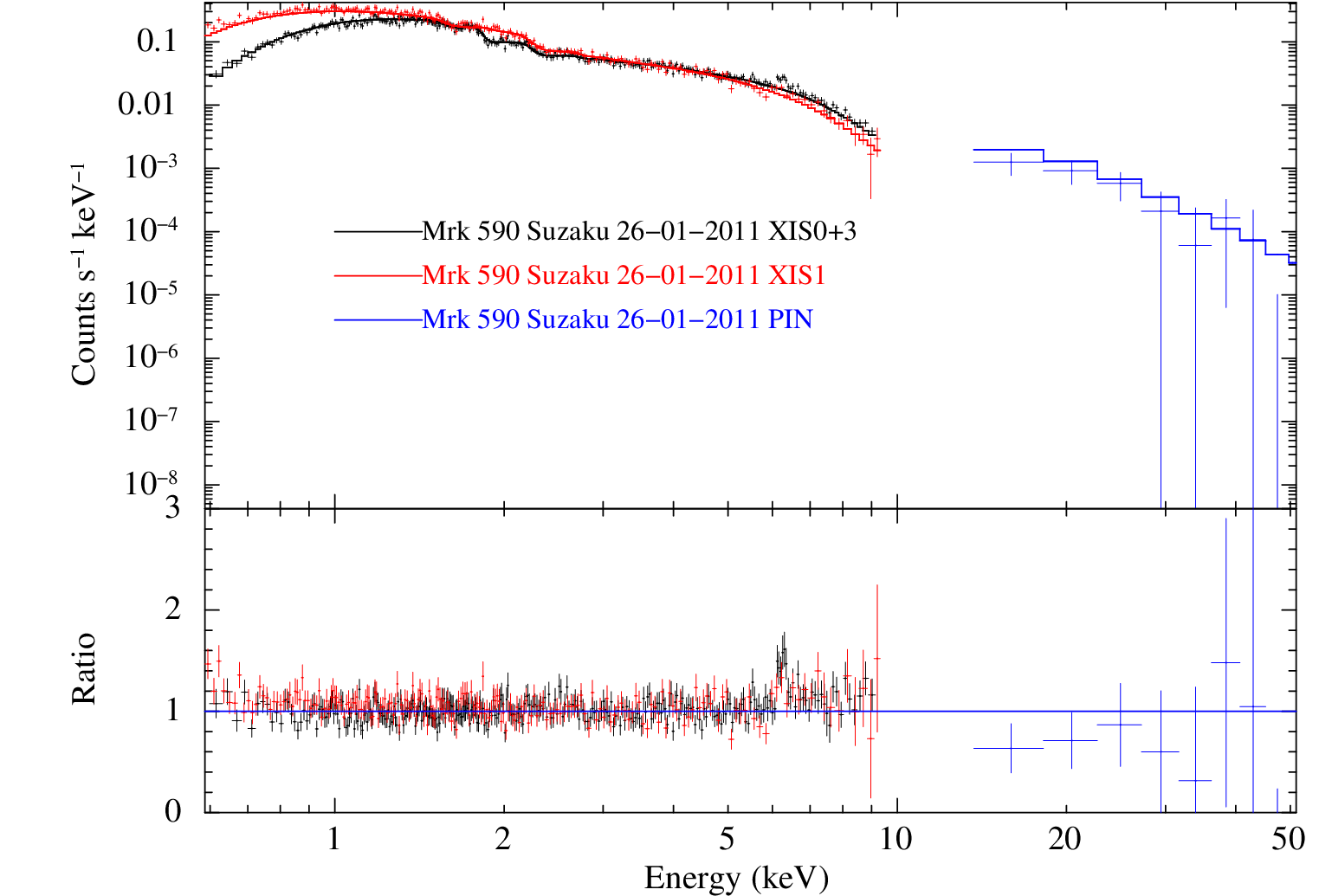}}
    \caption{{\it Left:}  The $2.0-5.0\kev$ \suzaku{} spectra, obs3 (on left) and obs4 (on right) of Mrk~590 fitted with an absorbed power-law and the rest of the energy band ($0.6-50.0\kev$) extrapolated. The broadband residuals from the fit, showing no soft X-ray excess and only Fe emission line complex  around $6.4\kev$ for the two \suzaku{} observations. The X-axis represents observed frame energy.} \label{fig:check_excess_obs3_4}

    \end{figure}

    \begin{figure}
    \centering 

    \hbox{
    \includegraphics[width=5.5cm,angle=-90]{check.excess.nustar.swift.5feb.2016.eps}
    \includegraphics[width=5.6cm,angle=-90]{check.excess.nustar.swift.27oct2018.eps}}
    \caption{{\it Left:}  The $2.0-5.0\kev$ \nustar{} and \swift{} spectra, obs5 (on left) and obs6 (on right) of Mrk~590 fitted with an absorbed power-law and the rest of the energy band ($0.6-50.0\kev$) extrapolated. The broadband residuals from the fit, showing no soft X-ray excess and only Fe emission line complex  around $6.4\kev$ similar to the two \suzaku{} observations. The X-axis represents observed frame energy.} \label{fig:check_excess_obs5_6}

    \end{figure}

    \begin{figure}
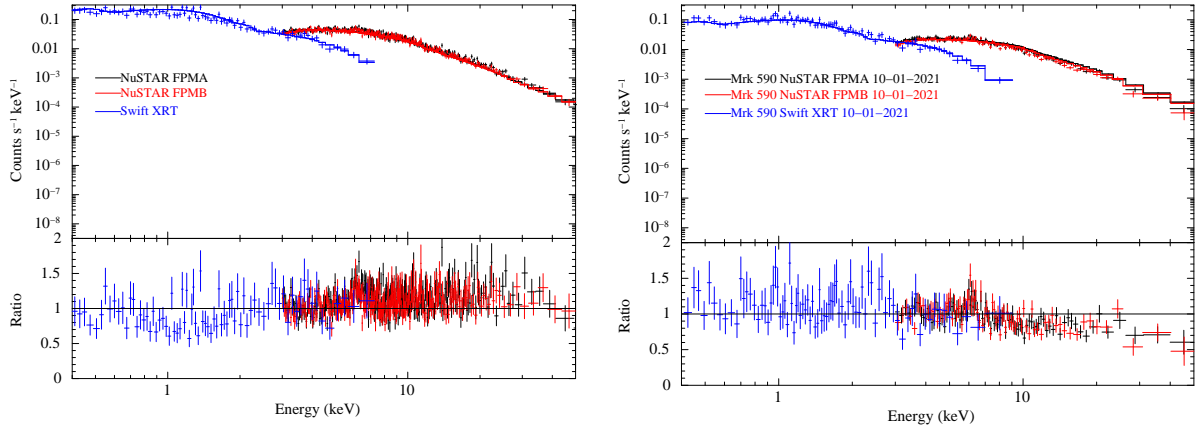

    \centering 
    \hbox{
    \includegraphics[width=5.5cm,angle=-90]{best_fit_check_excess_obs7.eps}
    \includegraphics[width=5.6cm,angle=-90]{best_fit_check_excess_obs8.eps}}
    \caption{Same as Fig.\ref{fig:check_excess_obs5_6} but for obs7 (on left) and obs8 (on right). The broadband residuals from the fit, showing no soft X-ray excess and only Fe emission line complex  around $6.4\kev$ similar to the two \suzaku{} observations. The X-axis represents observed frame energy.} \label{fig:check_excess_obs7_8}
    \end{figure}

\clearpage

\section{The spectral fit of Mrk 590 with the model relxill plus MyTorus at different epochs}\label{appendix_relxill}

\clearpage

\begin{figure}
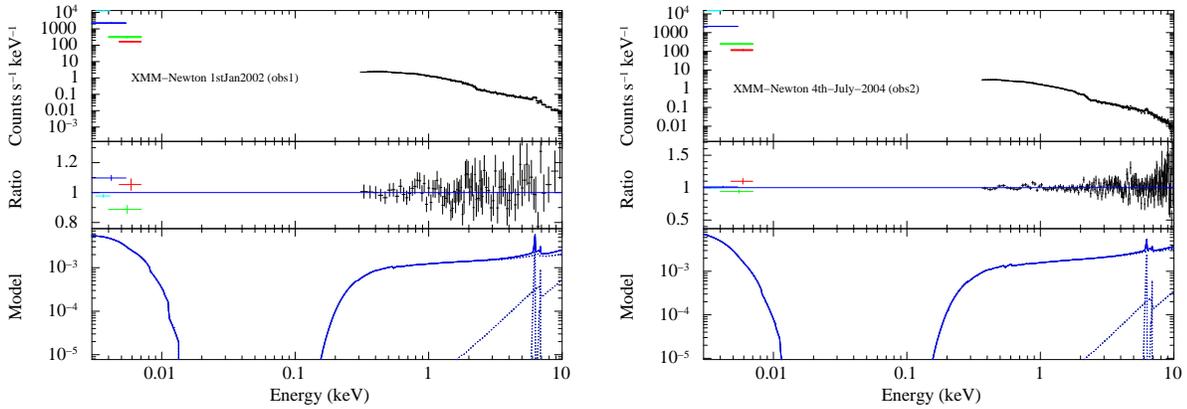

  \centering 
\hbox{
\includegraphics[width=5.5cm,angle=-90]{Best.fit.abs.relxill.mytorus.diskbb.withoutspin_withOM_obs1.eps}
\includegraphics[width=5.5cm,angle=-90]{Best.fit.abs.relxill.mytorus.withoutspin.diskbb.OM_obs2.eps}}
\caption{ The $0.001-10.0\kev$ \xmm{} EPIC-pn and OM data of Mrk~590 fitted with an absorbed {\it relxill} and {\it MyTorus} model. The data, the residuals and the theoretical model shown for obs1 (Left) and obs2 (Right). The X-axis represents observed frame energy.} \label{fig:relxill_fit_obs1_2}
\end{figure}

\begin{figure}
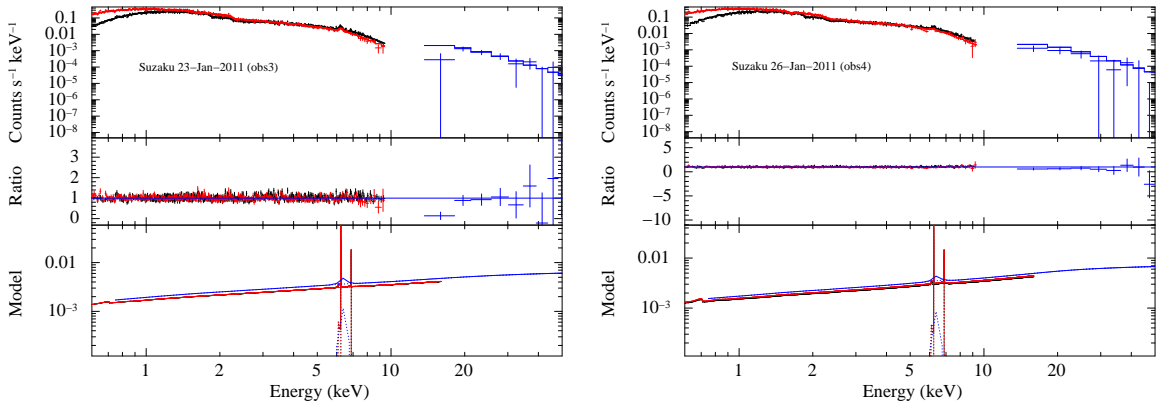

  \centering 
\hbox{
\includegraphics[width=5.5cm,angle=-90]{Best.fit.abs.relxill.mytorus.withoutspin.Ecut_fixed.suzaku.62ks_obs3.eps}
\includegraphics[width=5.5cm,angle=-90]{Best.fit.abs.relxill.mytorus.withoutspin.Ecut_fixed.suzaku.41ks_obs4.eps}}
\caption{The $0.6-50.0\kev$ \suzaku{} XIS and PIN data of Mrk~590 fitted with an absorbed {\it relxill} and {\it MyTorus} model. The data, the residuals and the theoretical model shown for obs3 (Left) and obs4 (Right). The X-axis represents observed frame energy.} \label{fig:relxill_fit_obs3_4}
\end{figure}

\begin{figure}
  \centering 
\hbox{
\includegraphics[width=5.5cm,angle=-90]{Best.fit.abs.relxill.mytorus.diskbb.withoutspin.swift.nustar.OM.5feb2016_obs5.eps}
\includegraphics[width=5.6cm,angle=-90]{Best.fit.abs.relxill.mytorus.withoutspin.diskbb.swift.nustar.UVOT.27oct2018_obs6.eps}}
\caption{The $0.001-10.0\kev$ \nustar{} FPM and \swift{} XRT and UVOT data of Mrk~590 fitted with an absorbed {\it relxill} and {\it MyTorus} model. The data, the residuals and the theoretical model shown for obs5 (Left) and obs6 (Right). The X-axis represents observed frame energy.} \label{fig:relxill_fit_obs5_6}
\end{figure}

\begin{figure}
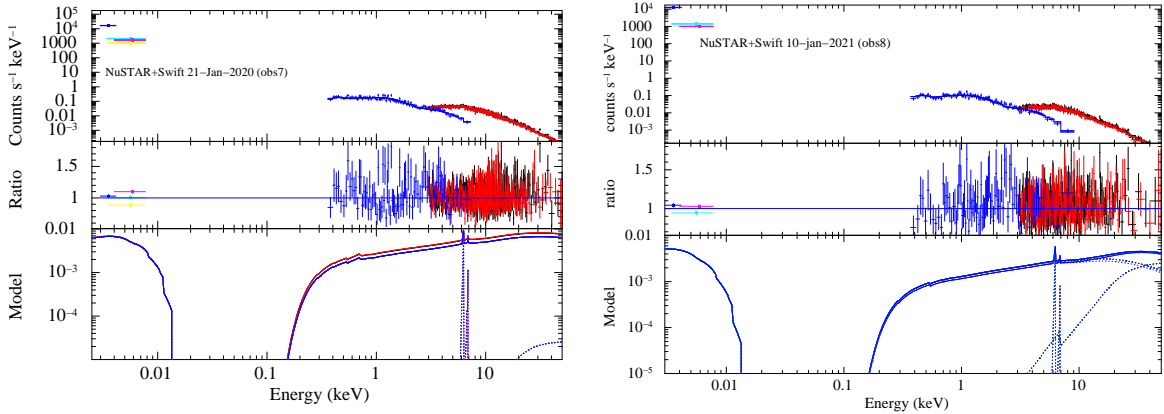

  \centering 
\hbox{
\includegraphics[width=5.5cm,angle=-90]{best_fit_abs_relxill_mytorus_withoutspin_uvot_obs7.eps}
\includegraphics[width=5.6cm,angle=-90]{best_fit_abs_relxill_withoutspin_mytorus_new_uvot_obs8.eps}}
\caption{Same as Fig.~\ref{fig:relxill_fit_obs5_6} but for obs7 (Left) and obs8 (Right). The X-axis represents observed frame energy.}\label{fig:relxill_fit_obs7_8}

\end{figure}

\bibliographystyle{aasjournal}
\bibliography{main}

\end{document}